\documentclass {article}

\usepackage{sprocl}

\usepackage{graphics}

\begin{document}

\title{
HADROPRODUCTION OF CHARM AND BEAUTY
\footnote{to appear in Proceedings of the XVI International Conference on
Physics in Collision, Mexico City, 1996.}
\\}

\author{PAUL E. KARCHIN }

\address{
Wayne State University,
Department of Physics and Astronomy, \\
666 West Hancock Street, 
Detroit, Michigan  48201, USA
}

\maketitle\abstracts{ Recent results on the hadroproduction of charm
  and beauty are discussed in the context of the current theoretical
  and experimental status.  The areas covered include production of
  open charm and beauty particles and charm and beauty quarkonia.
  Experimental results from both fixed target and colliding beam
  experiments are presented and compared to theoretical predictions.
  Predictions based on QCD perturbation theory are generally
  successful in explaining the shapes of differential cross sections
  but underestimate the scale of the cross section by factors of 2-3.
  An exception is production of the vector meson quarkonia which
  cannot be satisfactorily explained by perturbative processes alone.
  }

\section{Introduction}

This review emphasizes recent progress in the understanding of the
hadroproduction of charm and beauty with an attempt to explain the new
developments in the context of the fundamental questions of interest.
Experimental and theoretical work is presented which best illustrates
the subject. In cases where two or more illustrations of the same
topic are available, usually one is chosen for presentation here and
the others cited as references. In making these choices, some
preference is given to results published in the regular journals over
preliminary results. However, important preliminary results are also
presented here. As there are many recent publications in this field,
the author sincerely hopes that the fraction of the available results
presented here serves to stimulate interest in the whole. 

\section{Theory}

\begin{figure}[hbtp] 
\begin{center}
  \resizebox{\width}{\height}{\includegraphics*[221,537][390,741]
    {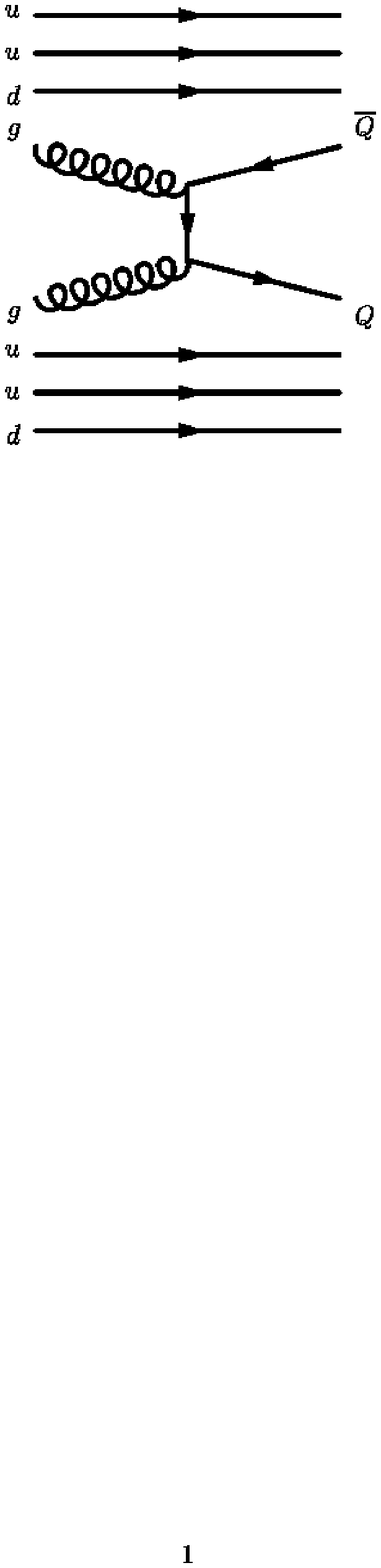}} {\caption{\label{fig:pp_collision} \protect
      \small Feynman diagram illustrating a $pp$ collision producing a
      heavy quark pair .  }}
\end{center}
\end{figure}

\begin{figure}[hbtp] 
%  \vspace{6cm} 
\begin{center}
  \resizebox{\width}{\height}{\includegraphics*[246,574][366,677]
    {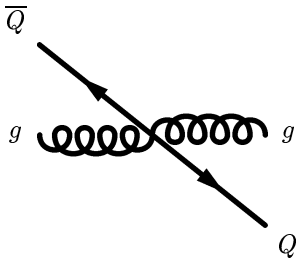}} {\caption{\label{fig:gg_collision} \protect \small
      The gluon gluon collision in its center of mass frame.  }}
\end{center}
\end{figure}

At high energy, heavy quark production is dominated by gluon
collisions. For example, in a $pp$ collision, illustrated in Figure
\ref{fig:pp_collision}, the valence quarks are spectators while the
hard interaction producing the heavy quark pair is via gluon fusion.
Many important characteristics of the heavy quark production
properties are evident from consideration of the gluon-gluon collision
in its center of mass frame as illustrated in Figure
\ref{fig:gg_collision}. At high energy, the $g-g$ $CM$ energy is given by
\begin{equation}
\hat{s}=x_1 x_2 s
\end{equation}
where $x_i$ are the momentum fractions of the gluons and $s$ is the
hadron-hadron $CM$ energy. Since $\hat{s}$ must be above
threshold for a pair of heavy quarks, the cross section clearly
depends on the gluon distribution functions of the colliding hadrons.
Thus, pions, kaons and protons incident on the same target will produce
different heavy quark cross sections. The $g-g ~CM$ frame is in
general moving with respect to the hadron $CM$ frame, and the heavy quarks
can have substantial longitudinal boosts in the hadron $CM$ frame.
The more unequal the gluon momentum fractions, the larger will be this
longitudinal boost. Since the gluon distribution functions peak at
small $x$, the heavy quark longitudinal momentum 
distributions (in the hadron-hadron $CM$) will peak at zero
and fall rapidly with increasing magnitude of the longitudinal momentum.

The $HQ$ production cross section is given by \cite{Nason},
\begin{equation}
\sigma = \int dx_1 dx_2 \hat{\sigma}(\hat{s}, q^2,M_Q) 
g_1(x_1,q_1^2)g_2(x_2,q_2^2)
\end{equation}
where $g_i(x_i,q_i^2)$ are the gluon distribution functions for
momentum fraction $x_i$ and momentum transfer $q_i^2$, and
$\hat{\sigma}$ is the short distance cross section.  The strong
coupling constant appears in $\hat{\sigma}$ and has the form,
\begin{equation}
\alpha_s(q^2) \sim { 1 \over ln{q^2 \over \Lambda_{QCD}^2}}.
\end{equation}
Perturbation theory can be used to evaluate $\hat{\sigma}$ when 
the coupling constant is less than one, as occurs when $q^2$,
which is of order $M_Q^2$, is greater than $\Lambda_{QCD}^2$.
This is the case for the charm and beauty quarks.

However, there are a number of theoretical problems with the formalism of
equation {2}. 
In existing calculations, the gluon distribution functions
do not include heavy quarks in their evolution. Second, there are terms
in the short distance cross section with coupling constants that become
larger than one 
in certain kinematic regions.  For example, factors of the form,
\begin{equation}
\alpha_s(q^2) ln {s \over q^2}
\end{equation}
become large for $s > q^2$.  For existing calculations, 
the theoretical problems are manifested by large sensitivity to changes in
the renormalization and factorization scales, as discussed later.

\begin{figure}[hbtp] 
%  \vspace{6cm} 
\begin{center}
  \resizebox{!}{12cm}{\includegraphics*[180,312][430,742]
    {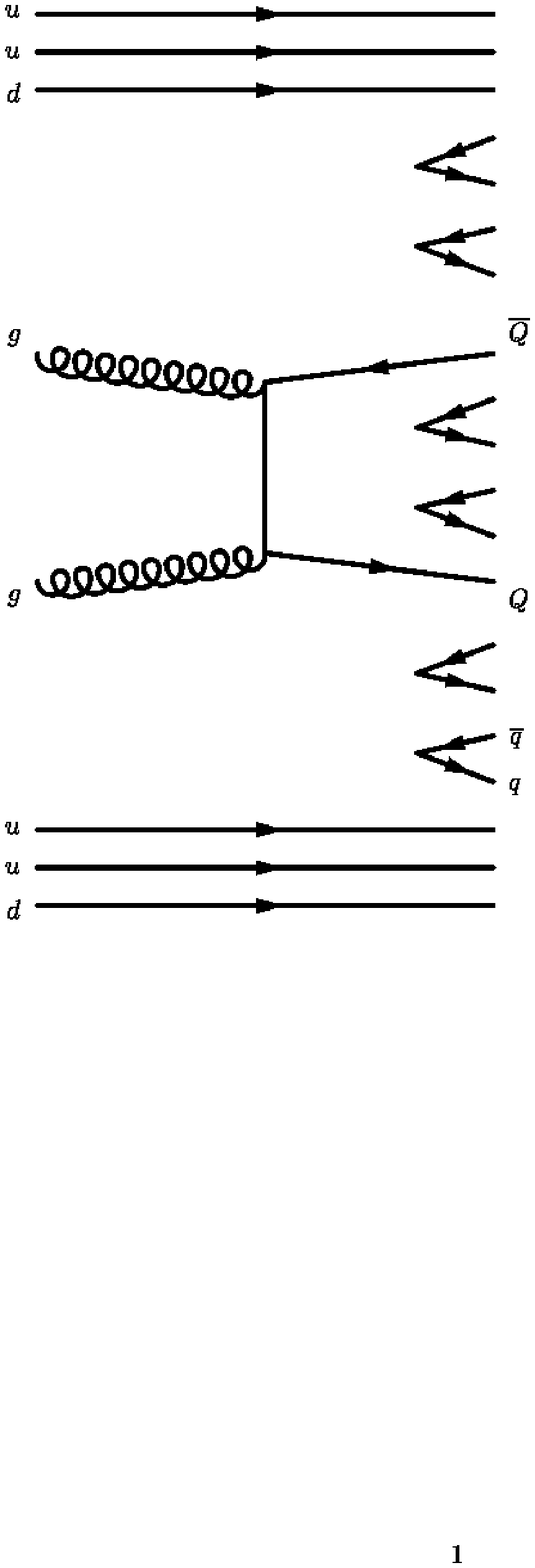}} {\caption{\label{fig:hadronization} \protect
      \small The hadronization (also called fragmentation) process.
      Light quark pairs can form ``strings'' between the heavy quarks
      and between a heavy quark and a spectator quark.  }}
\end{center}
\end{figure}

If the produced heavy quarks are close enough in momentum, they can
form a bound state: the quarkonia states of charm or beauty. This
formation process has a small probability and hence the fraction of
heavy quark production into quarkonia is small, of order 1 part in
1000.  The bulk of heavy quark production goes into open charm or
beauty states.  These cross sections depend on the formation of heavy
quarks into particles, a process often referred to as hadronization or
fragmentation. Several sub-processes can contribute to this hadronization,
as illustrated in Figure \ref{fig:hadronization}.  The heavy quarks
can interact with light quark/ anti-quark pairs arising from 
virtual gluons.  Also, the heavy quarks can interact with light
valence quarks from the colliding hadrons which are spectators to the
hard interaction which produced the heavy quark pair.
The hadronization process can involve many small $q^2$
interactions and hence is intrinsically non-perturbative.
Phenomenological models, such as the Lund string model, have been
developed to describe the hadronization process.

Because the amount of heavy quark fragmentation into a given charm
species cannot be reliably treated theoretically, comparison of the
predicted total heavy quark cross section with experiment must include
measurement of as many species (mesons and baryons) as possible.
Comparison of measured differential cross sections for particles with
theoretical predictions for quarks can reveal the effect of
fragmentation.

As a preliminary remark, we note that rapidity, $y$, and Feynman-$x$, $x_F$, 
are commonly used as variables to measure the
longitudinal momentum of the produced heavy particle. 
Rapidity is the standard variable of special relativity.
Feynman-$x$ is defined, in the $CM$ frame of the colliding hadrons,
as the ratio of the 3-momentum of the heavy particle (or quark) to the maximum
possible 3-momentum.  Hence, the range of $x_F$ is from -1 to 1.

\section{Open Charm}

\begin{figure}[hbtp] 

  \resizebox{\textwidth}{!}{\includegraphics*[16,150][525,650]{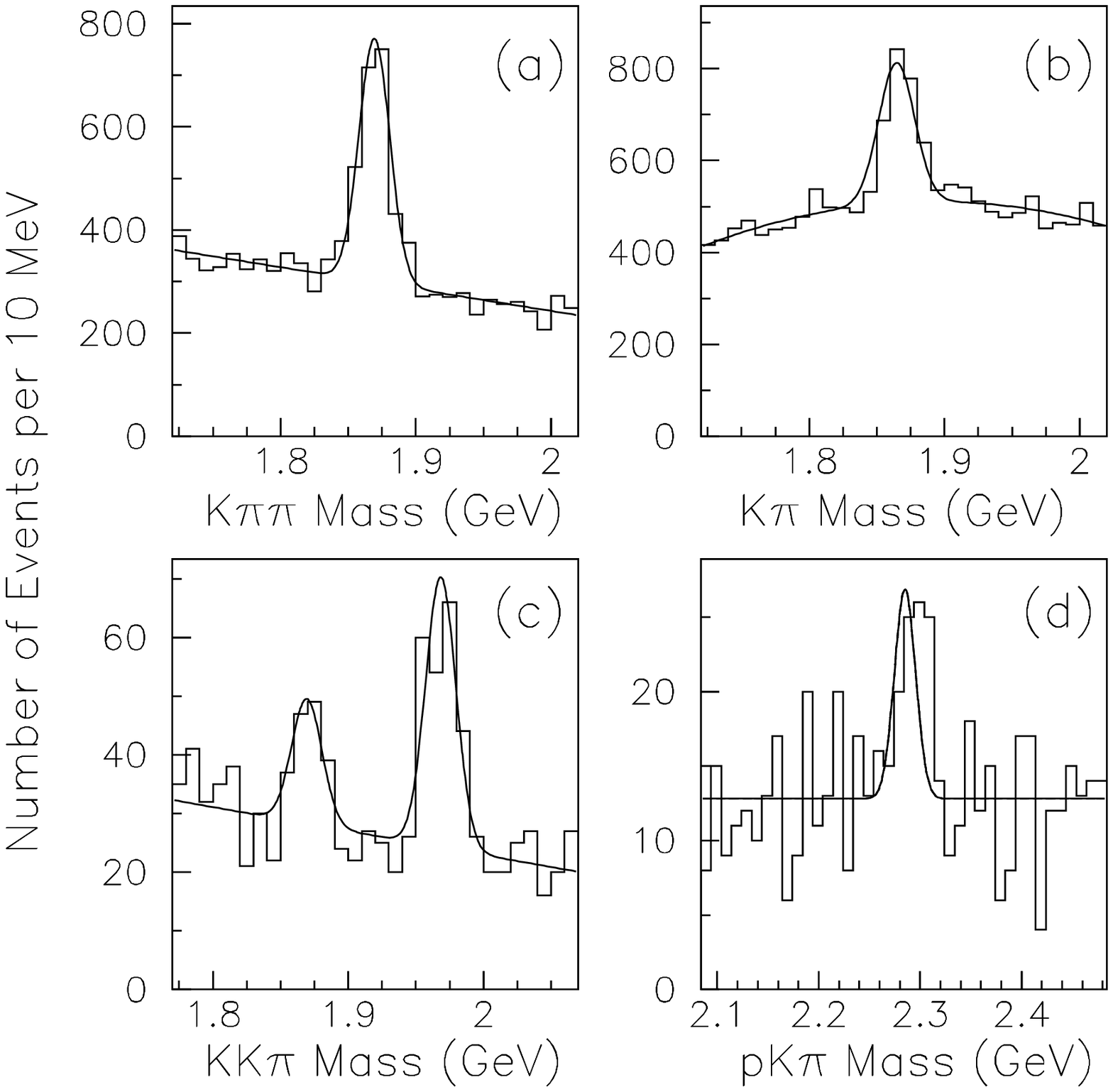}}
  {\caption{\label{fig:E769_mass_plots} (from ref. 2)
      Invariant mass distributions from Fermilab E769.  The decay
      modes (including charge conjugates) and number of signal events
      with statistical errors are (a) $D^{+} \rightarrow K^{-} \pi^{+}
      \pi^{+}$, $994\pm47$, (b) $D^{0} \rightarrow K^{-} \pi^{+}$,
      $847\pm55$, (c) $D_{s}^{+} \rightarrow K^{+} K^{-} \pi^+$
      (combined $\phi$ $\pi$ and $K^*K$ modes), $100\pm15$, (d)
      $\Lambda_c^+ \rightarrow p K^- \pi^+$, $35\pm9$. The signals
      shown are combined from all beam types.  }}
\end{figure}

There are new measurements of the production of open charm states from
Fermilab Experiment E769 \cite{E769A,E769B}.  
This experiment utilized a 250 GeV 
charged secondary beam with the identified beam particles $\pi^+$,
$\pi^-$, $K^+$, $K^-$, and $p$.  The charm particles 
$D^+$, $D^0$, $D^{*+}$, $D_S$
and $\Lambda_c$ are measured using fully reconstructed decays to all charged
final states.  Examples of mass plots are shown in 
Figure \ref{fig:E769_mass_plots}.  The
signal to background ratios and signal sizes are sufficient to measure
the total and differential cross sections for 
most of the charm particles and beam types mentioned above.  In some
cases, only upper limits on cross sections are possible.

%\epsfbox{fig1.ps} 
%\vspace{5in}
%\special{psfile=fig1.ps hoffset=-100 voffset=-150 hsize=550 vsize=520
%hscale =80 vscale=80}
%\includegraphics*[80,150][660,700]{fig1.ps}
%\includegraphics[bb=80,150,660,700, width =\textwidth]{fig1.ps}

E769 is a fixed target experiment utilizing a forward spectrometer.
The acceptance for reconstructed charm particle decays is typically
over the range $-.1 < x_F < .9$.  Charm particle cross
sections are reported for the $x_F > 0$, a range convenient for
comparison with other experiments.  Also convenient for
comparing different experiments and theory is the sum of the inclusive
cross sections,
\begin{equation}\label{eqn:summed_sigma}
\sigma(D^+) + \sigma(D^-) + \sigma(D^0) + \sigma(\overline{D}^0) 
+\sigma(D_s^+) + \sigma(D_s^-)
\end{equation}  
for $x_F > 0$.  This sum provides a partial measure of the total charm
cross section $\sigma_{c\overline{c}}$ for producing a pair of charm and
anti-charm quarks. 

\begin{figure}[hbtp] 

  \resizebox{\textwidth}{!}{\includegraphics*[16,150][525,650]{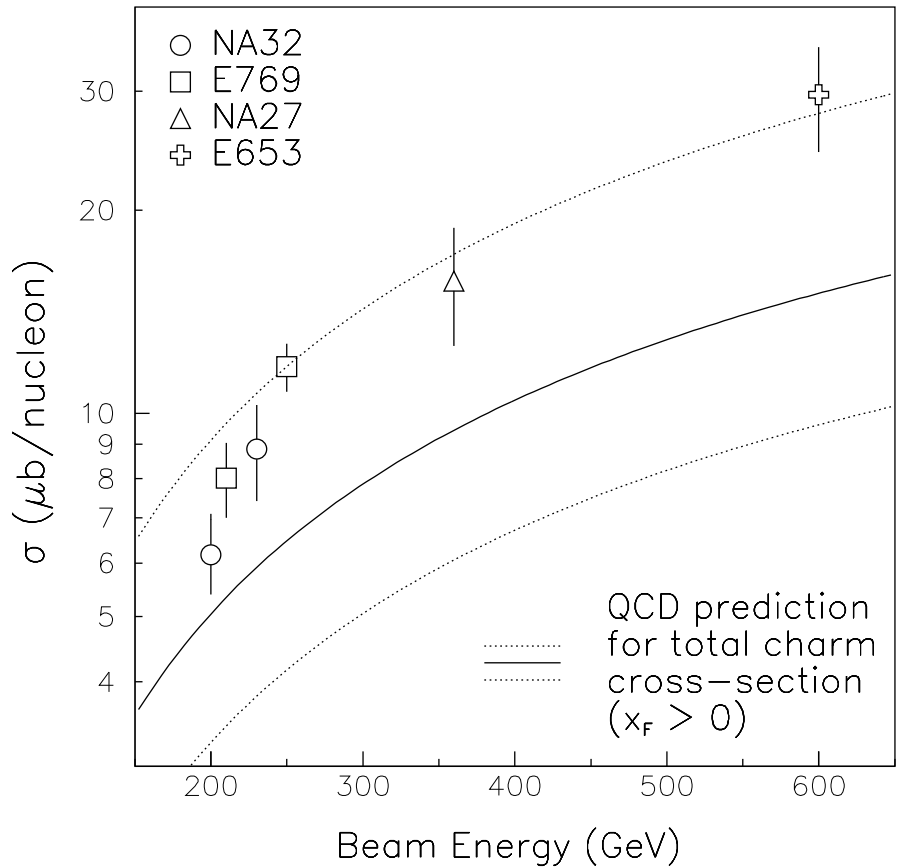}}
  {\caption{\label{fig:pion_xsec} \protect \small (from
      ref. 2) Measurements of the summed cross section in
      equation (\ref{eqn:summed_sigma}) for $\pi^-$ nucleon
      interactions from E769 and other experiments plotted versus beam
      energy.  The NLO QCD predictions are for the sum of the charm
      quark and anti-charm quark cross section for $x_F>0$. The upper
      and lower curves result from variation of the renormalization
      scale from $m_c$ down to $m_c/2$ and up to $2m_c$, respectively.  }}
\end{figure}

Measurements of summed cross sections in equation 
(\ref{eqn:summed_sigma}) for
$\pi^-$-nucleon interactions from E769 and other experiments are
plotted versus beam energy in Figure \ref{fig:pion_xsec}.  Also shown
are theoretical predictions based on perturbative QCD for different
values of the renormalization scale, $\mu_R$.  The predictions are for
twice the cross section for producing a charm quark pair for which at
least one of the pair has $x_F > 0$. These predictions are generated
using the computer program of Mangano, Nason and Ridolfi.  It is clear
from Figure \ref{fig:pion_xsec} that the measured shape of the energy
dependence is in reasonable agreement with theory and that the scale
of the measured cross sections are higher than the central value of
the prediction.  Keep in mind that the summed cross section does not
include charm baryons, which would increase the discrepancy if
included.  As can be seen from the curves for different values of
$\mu_R$, the prediction has significant variation.  This is also the
case for variation of the factorization scale.

\begin{figure}[hbtp] 

  \resizebox{\textwidth}{!}{\includegraphics*[16,150][525,650]{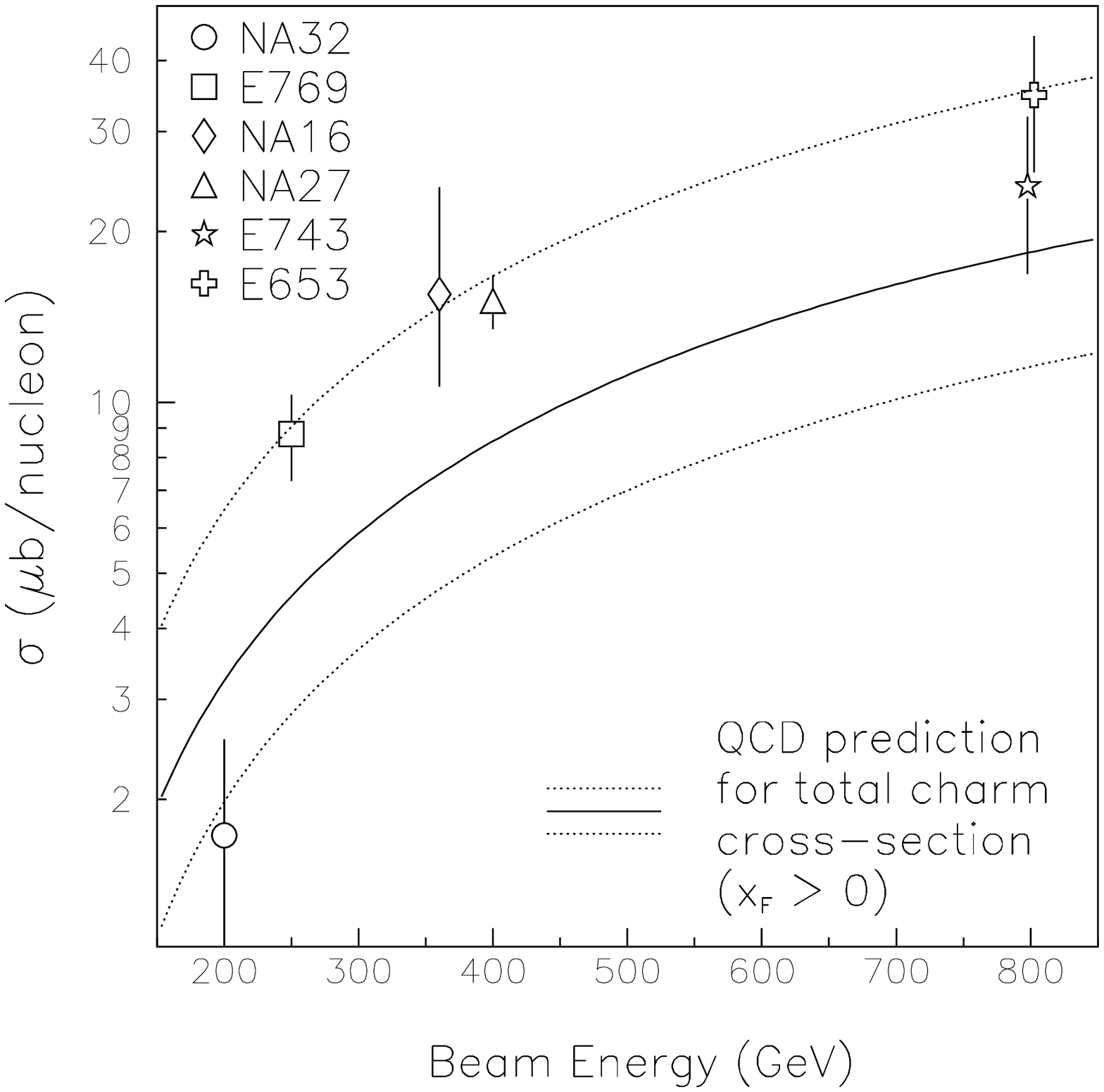}}
  {\caption{\label{fig:proton_xsec} \protect \small (from
      ref. 2) Measurements of the summed cross section in
      equation (\ref{eqn:summed_sigma}) for $p$ nucleon interactions
      from E769 and other experiments plotted versus beam energy.  The
      NLO QCD predictions are for the sum of the charm quark and anti-
      charm quark cross section for $x_F>0$.  The upper and lower
      curves result from variation of the renormalization scale from
      $m_c$ down to $m_c/2$ and up to $2m_c$, respectively.  }}
\end{figure}

For $p$ nucleon interactions, the comparison of the summed cross
section with theory is shown in Figure \ref{fig:proton_xsec}.  The
comparison shows the same qualitative behavior as for the 
$\pi^-$-nucleon case.

There are no theoretical predictions for the kaon- nucleon cross
sections since no parton distribution functions are available for this
case.  Cross sections for five different charm species and for pion,
kaon and proton beams, measured by E769, are compared in Table 
\ref{tab:E769_xsec}.
Generally, for a given charm species, the cross sections are similar
for the different beams.  One notable exception appears to be the
relatively large cross section for $\Lambda_c$ production with protons
as compared to pions and kaons.  This effect may indicate that an
initial state light baryon enhances the production of a charm baryon.
An analogous test is whether an initial state strange particle (kaon)
enhances the production of charmed strange mesons $(D_s^{\pm})$. From
the table, one sees 
no indication of this effect: 
the kaon induced $D_s$ cross section is consistent with 
the $D_s$ cross sections for
pions and protons.

The relative size of the $\Lambda_c$ and $D_s$ cross sections to those 
of the light charmed mesons is important for 
measuring the 
total charm cross section.  
The data of Table \ref{tab:E769_xsec} from E769 indicate that 
the cross sections for 
$\Lambda_c$ and $D_s$ production are comparable to those of $D^+$ and $D^0$.
This is also supported by earlier data from CERN experiment
NA32 (ACCMOR)\cite{ACCMOR} which are compared to those from E769 
in Table \ref{tab:compare_NA32}. 
Thus, it appears that all charm species have comparable 
cross sections.  The summed cross sections for all charm particles is
at least a factor of three more than the central NLO QCD prediction.

\begin{table}[hbtp]
\caption{Charm particle (including anti particle) cross sections 
in $\mu$b/nucleon for $x_F > 0$ measured by E769. 
\label{tab:E769_xsec}}
\vspace{0.4cm}
\begin{center}
\begin{tabular}{|l|c|c|c|}
\hline
            & $\pi^{\pm}$     & $K^{\pm}$ & $p$ \\
\hline
$D^+$       & $3.2\pm0.2\pm0.2$ & $3.0\pm0.3\pm0.2$ & $3.2\pm0.4\pm0.3$ \\ 
$D^0$       & $7.2\pm0.5\pm0.4$ & $7.2\pm1.0\pm0.4$ & $5.6\pm1.3\pm0.5$ \\ 
$D_s$       & $2.0\pm0.4\pm0.2$ & $3.0\pm0.8\pm0.3$ & $>0.5,<2.5$       \\ 
$D^{*+}$    & $2.8\pm0.3\pm0.2$ & $1.7\pm0.5\pm0.1$ & $1.8\pm0.6\pm0.1$ \\ 
$\Lambda_c$ & $3.3\pm1.1\pm0.5$ & $1.8\pm0.6\pm0.1$ & $>5.0,<21.2$      \\ 
\hline
\end{tabular}
\end{center}
\end{table}

\begin{table}[hbtp]
\caption{Comparison of cross sections 
(including anti- particle)
in $\mu$b/nucleon for $x_F > 0$ measured by 
NA32 (ACCMOR, 230 GeV $\pi^-$ beam)
and
E769 (250 GeV $\pi^{\pm}$ beam ).
\label{tab:compare_NA32}}
\vspace{0.4cm}
\begin{center}
\begin{tabular}{|l|c|c|}
\hline
            & NA32 & E769 \\
\hline
$D_s$       & $1.4\pm0.2\pm0.2$ & $2.0\pm0.4\pm0.2$ \\ 
$\Lambda_c$ & $4.1\pm0.5\pm0.7$ & $3.3\pm1.1\pm0.5$ \\ 
\hline
\end{tabular}
\end{center}
\end{table}

Another new result from E769 is the measurement of differential charm
cross sections in $x_F$ and $P_T$ for pion, kaon and proton beams
\cite{E769B}.  As we will see, the cross sections are affected by the
parton distributions in the different beam particles.

\begin{figure}[hbtp] 

  \resizebox{\textwidth}{!}{\includegraphics*[16,396][545,670]{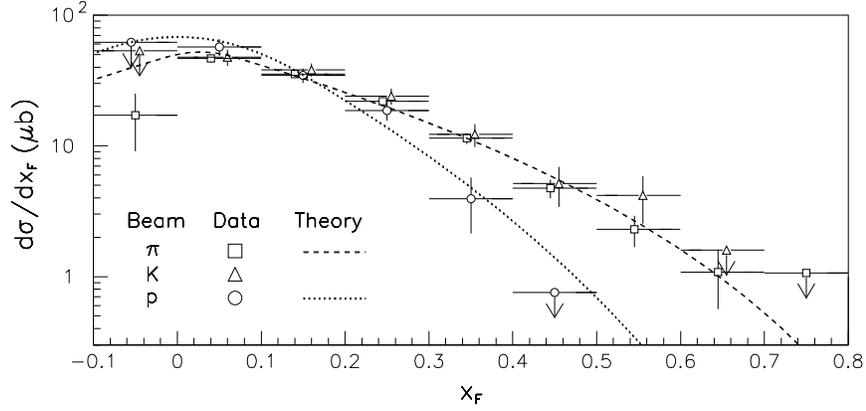}}
  {\caption{\label{fig:xf_prl} \protect \small (from ref. 3)
      E769 measurements of $d\sigma/dx_F$ for charmed mesons produced
      with $\pi$, $K$ and $p$ beams.  Only statistical errors are
      shown.  Arrows indicate 90\% confidence level upper limits.  The
      predictions are for charm quarks, based on NLO QCD.  }}
\end{figure}

For the E769 measurements, in each bin of the variable of interest
($x_F$ or $P_T$), the cross section is summed as in equation
(\ref{eqn:summed_sigma}). (The summation increases the statistical power
of the measurement.) The measurements for $d\sigma/dx_F$ are shown in
Figure \ref{fig:xf_prl}. Also shown are the NLO QCD predictions for
charm quark production from pion and proton beams.  (The predictions
are normalized to best fit the data; the shape of the prediction has
no constraints from the data.) It is clear that the expected shape for
the proton beam is steeper in $x_F$ than for the pion beam.  Indeed,
the $\pi$ and $p$ data are not consistent with having the same shape (C.L.
$>$ 99\%). Furthermore, the $\pi$ and $p$ data are well fit by the
predicted shapes (C.L. $>$ 50\%).  Thus, it appears that the $x_F$
dependence shows the expected effect of the beam parton distribution.

Although there is no prediction with a kaon beam, the data for
pion and kaon beams are consistent with having the same shape (C.L. $>$ 95\%).
This indicates that the pion and kaon have similar parton distributions.

\begin{figure}[hbtp] 

  \resizebox{\textwidth}{!}{\includegraphics*[16,396][545,670]{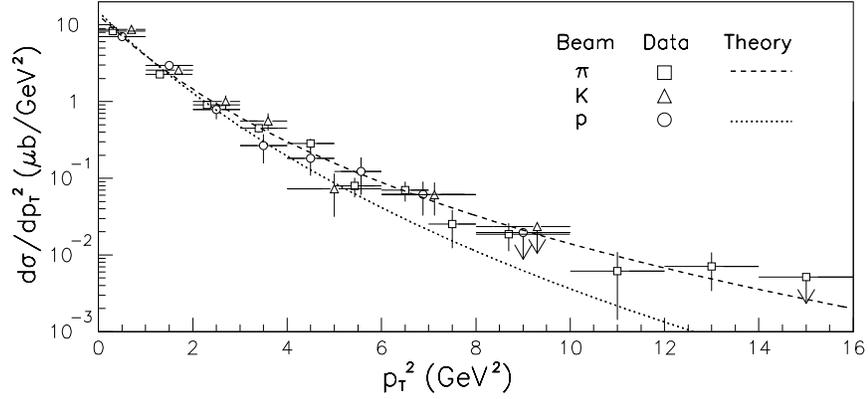}}
  {\caption{\label{fig:pt2_prl} \protect \small (from
      ref. 3) E769 measurements of $d\sigma/dP_T^2$ for
      charmed mesons with $x_F>0$ for charmed mesons produced with
      $\pi$, $K$ and $p$ beams. The predictions are for charm quarks,
      based on NLO QCD.  }}
\end{figure}

The differential cross sections in $P_T^2$ are shown in Figure
\ref{fig:pt2_prl}. The pion beam data are consistent with the NLO QCD
prediction for a pion beam, but inconsistent with the prediction for a
proton beam.  The difference in the $P_T$ distributions for pions and
protons is evident, although not as distinct as that for the $x_F$
distributions.

\begin{figure}[hbtp] 
%  \vspace{9cm} 
%Line below for scanned image.
  \resizebox{\textwidth}{!}{\includegraphics*[17,207][600,588]
    {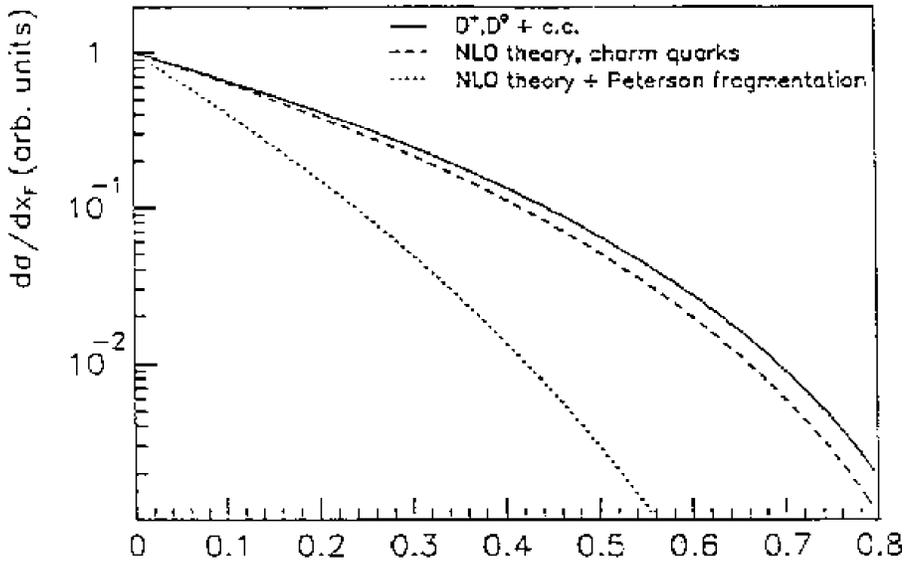}} {\caption{\label{fig:Wu} \protect \small The
      solid curve is a fit to an early measurement of the $D$ meson
      cross section versus $x_F$ from E769 with a pion beam.  The
      dashed curve is the NLO QCD prediction for charm quark
      production. The dotted curve results when the Peterson
      fragmentation function is convoluted with the NLO QCD prediction
      for quarks. }}
\end{figure}

One of the puzzles of charm particle production is that shapes of the
differential cross sections in $x_F$ and $P_T$ are so close to the NLO
QCD predictions for charm quarks.  This is certainly not the case in
$e^+e^-$ production, where the effect of fragmentation is substantial.
Indeed, if the Peterson fragmentation function is used to modify the
NLO QCD result for charm quark hadroproduction, the shape of the
prediction falls much too rapidly compared to the data as shown in
Figure \ref{fig:Wu}.  \nocite{Wu} Perhaps loss of charm
quark momentum from fragmentation is counterbalanced by an
acceleration from interaction with spectator quarks.

\section{Open Beauty}

In contrast to measurements of the charm cross section, most
measurements to date of the beauty cross section rely on partial
reconstruction techniques, as did charm measurements of a decade ago.
Many of these earlier charm measurements suffered from systematic
errors due to the assumed models of production and decay dynamics.
This history provides a lesson for the current measurements of beauty
production. For the measurements based on partial reconstruction
techniques, systematic errors in the beauty quark pair cross section are
difficult to avoid from model dependent assumptions about the
production and decay dynamics.

\subsection{Fixed Target Beauty Experiments}

\begin{figure}[hbtp] 

  \resizebox{\textwidth}{!}{\includegraphics*[36,46][583,639]{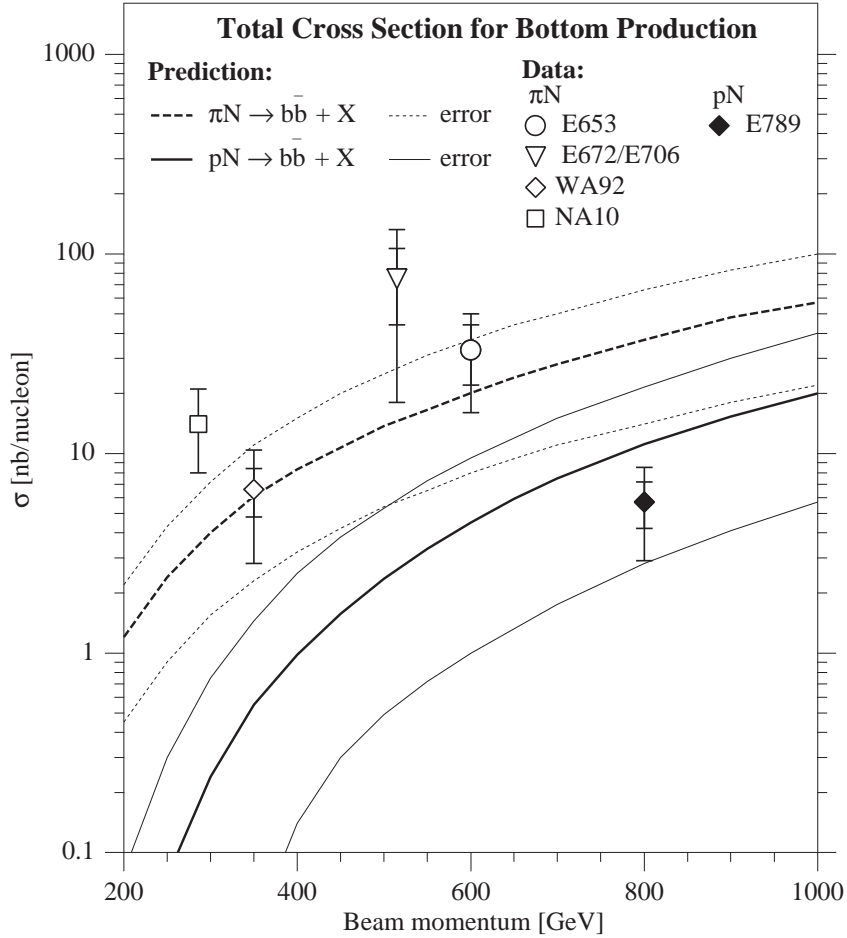}}
  {\caption{\label{fig:b_cross_ft} \protect \small Fixed target $B$
      cross section measurements. The solid lines are the predictions
      for a proton beam and the dotted lines are the predictions for a
      pion beam.  The predictions are based on NLO QCD (ref. 6).
      Theoretical curves are shown for central parameter values and
      variations of $\Lambda_{QCD} = 260\pm100 MeV$ and $\mu = \mu_0,
      \mu_0/2$, and $2\mu_0$.  }}
\end{figure}

A variety of partial reconstruction techniques have been employed to
measure the $b\overline{b}$ cross section in fixed target experiments.
The experiments and techniques employed are listed in Table
\ref{tab:ft_beauty_exper}.  The measurements, which are plotted in
Figure \ref{fig:b_cross_ft}, have large uncertainties, but generally
lie at or above the central values of the QCD prediction.
\nocite{Nason89}

\begin{table}[hbtp]
\caption{
Fixed target experiments measuring the beauty pair cross section.
\label{tab:ft_beauty_exper}
}
\vspace{0.4cm}
\begin{center}
\begin{tabular}{|l|l|}
\hline
Experiment Name                      & Technique        \\
\hline                                         
CERN NA10         & tri muon \\
FNAL E653         & nuclear emulsion \\
FNAL E672/E706    & inclusive $J/\psi$ with detached vertex \\
FNAL E789         & inclusive $J/\psi$ with detached vertex \\
CERN WA92         & silicon active target \\ 
\hline
\end{tabular}                                          
\end{center}                                           
\end{table}

\subsection{Measurements at the Tevatron}

\begin{figure}[hbtp] 
  \resizebox{\textwidth}{!}{\includegraphics*[63,328][299,532]
    {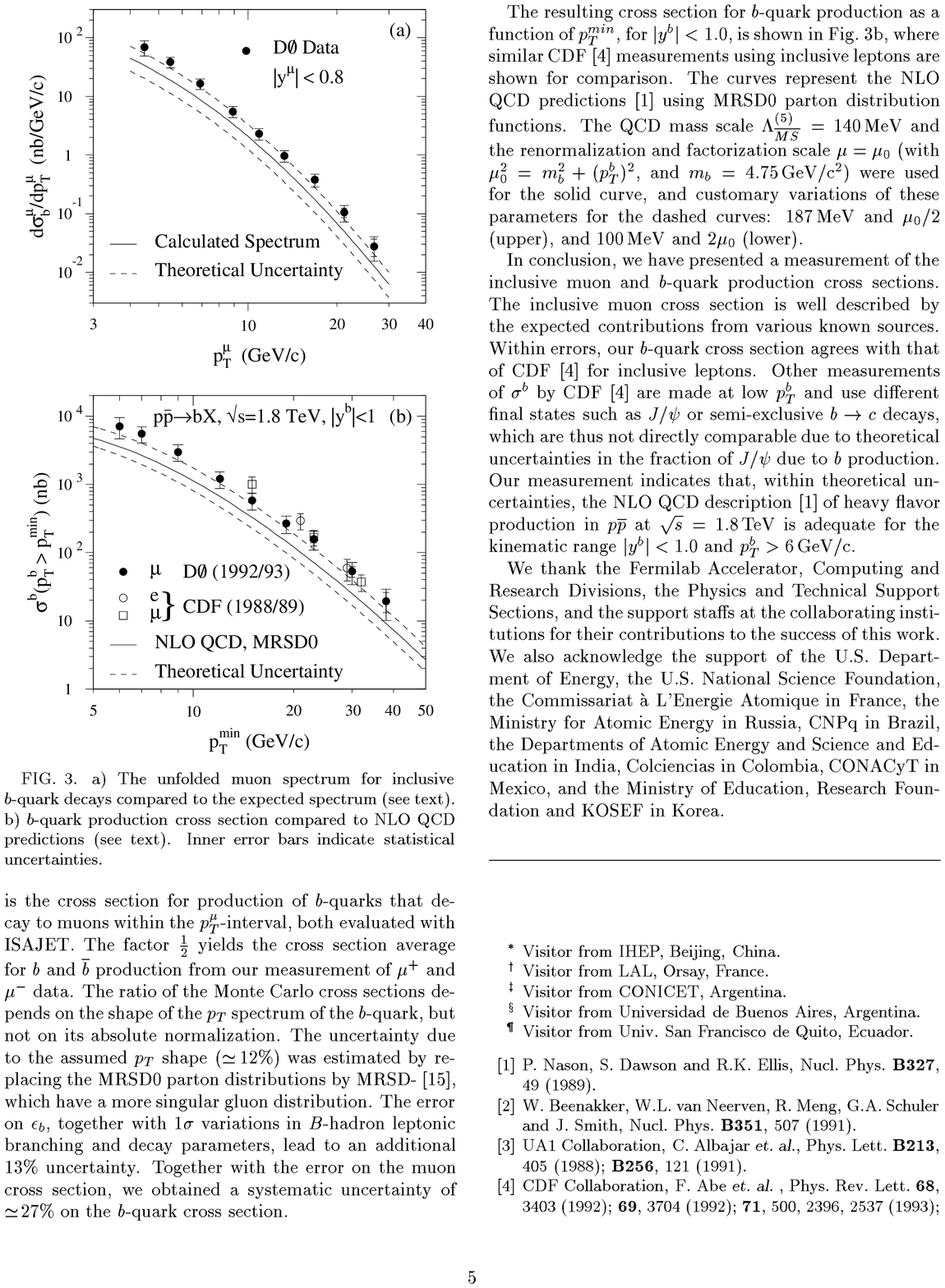} } {\caption{\label{fig:D0_incl_mu} \protect \small
      (from ref. 7) Measurements of the cross section for a $b$ quark
      derived from the spectrum of single leptons under the assumption
      of the ISAJET model. The solid and dotted lines show the
      prediction and uncertainty based on NLO QCD.  For the
      prediction, the central values of parameters are $\Lambda_{QCD}$
      = 140 MeV and $m_b$ = 4.75 GeV with $\mu_F$ = $\mu_R$ = $\mu_0$
      = $\sqrt{m_b^2 + (p_T^b)^2}$.  The curves showing the theory
      error are based on variations of ${\Lambda_{QCD}}^{+47}_{-40}$
      MeV, $\mu_0 \rightarrow \mu_0/2$ and $\mu_0 \rightarrow 2\mu_0$.
      }}
\end{figure}

At the Fermilab Tevatron collider, a well established technique to
measure the beauty cross section is via the inclusive cross section
for single high $P_t$ leptons. A recent measurement \cite{D0_muon} by
the D0 experiment, using muons, is shown in Figure
\ref{fig:D0_incl_mu} as well as earlier measurements from CDF.
Plotted is the cross section for a $b$ quark as a function of the
minimum $p_T$ of the $b$ quark. The plotted cross section is derived
under the assumption of the ISAJET model relating the $p_T$ of the
lepton to that of the parent $b$ quark.  The model takes into account
the momentum fragmentation of the $b$ quark, the speciation into
hadronic particles and the decay kinematics of those hadrons.  Since
few of these processes have been measured, systematic errors are
inherent. The $b$ quark cross sections extracted from both experiments
(using the same production model) are consistent with theory, but fall
along the upper range of the theoretical uncertainty.

\begin{figure}[hbtp] 
  \resizebox{\textwidth}{!}{\includegraphics*[75,268][487,661]
    {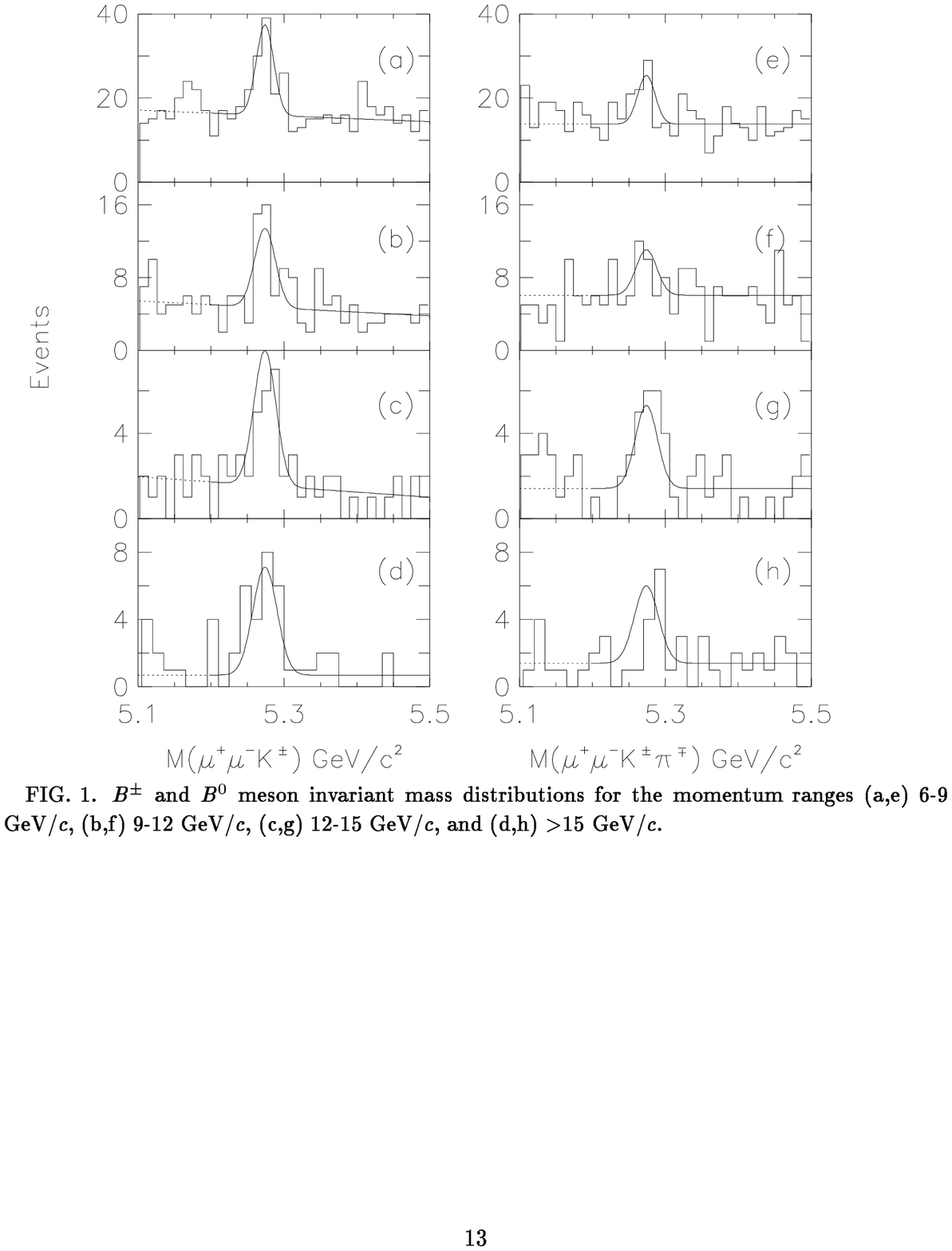} } {\caption{\label{fig:CDF_B_mass_plots}
      \protect \small (from ref. 8) Invariant mass plots from the CDF
      experiment for (left column) $B^+ \rightarrow \psi K^+$ and
      (right column) $B^0 \rightarrow \psi K^{*0}$ with $K^{*0}
      \rightarrow K^+\pi^-$.  The $P_T$ ranges are (a,e) 6-9 GeV/c,
      (b,f) 9-12 GeV/c, (c,g) 12-15 GeV/c and (d,h) $>15$ GeV/c.  The
      signal in (a) is $51\pm10$ events and in (e) is $72\pm12$
      events.  }}
\end{figure}

Beauty cross section measurements based on exclusive decay modes of
specific particles avoid some of the model dependence of the
measurements derived from inclusive lepton production.  The CDF
experiment \cite{CDF_B_meson} has measured the cross sections for the
production of $B^+$ and $B^0$ through the decay modes $B^+ \rightarrow
\psi K^+$ and $B^0 \rightarrow \psi K^{*0}$ with $K^{*0} \rightarrow
K^+\pi^-$.  Invariant mass plots for these decays, for different
momentum ranges of the parent $B$, are shown in Figure
\ref{fig:CDF_B_mass_plots}.  Analysis cuts include those based on the
$J/\psi$ and $K^*$ masses and proper lifetime greater than $100 \mu m$.

\begin{figure}[hbtp] 
  \resizebox{\textwidth}{!}{\includegraphics*[40,300][510,745]
    {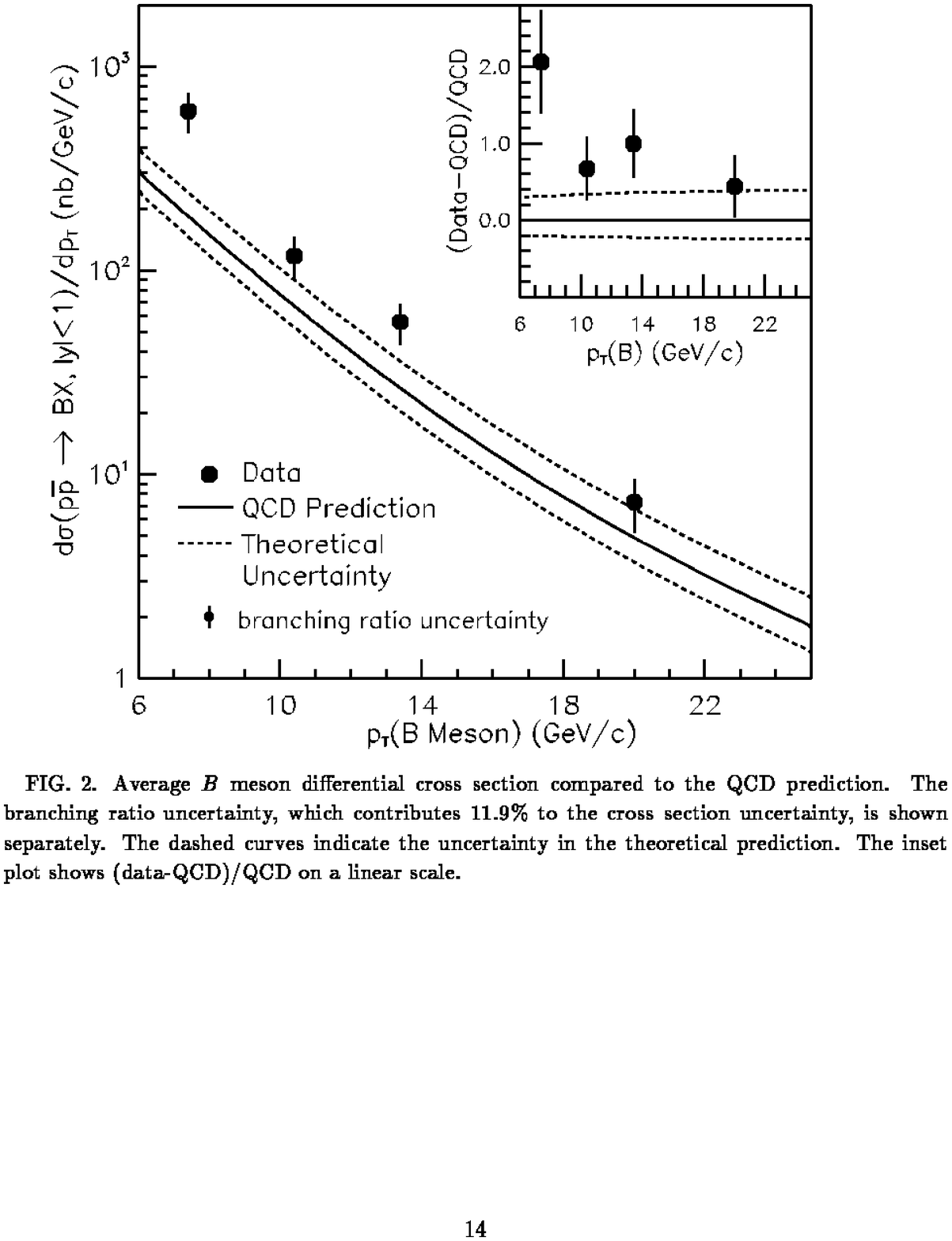} } {\caption{\label{fig:CDF_B_xsec} \protect \small
      (from ref. 8) CDF measurement of the average $B$ meson
      differential cross section based on the signals in the previous
      Figure.  The branching ratio uncertainty, which contributes
      11.9\% to the cross section uncertainty, is shown separately.
      The dashed curves indicate the uncertainty in the theoretical
      prediction due to the uncertainties in $m_b$, momentum
      fragmentation and renormalization scale.  The theoretical
      prediction assumes $f_{B^+} = f_{B^0} = .375$ for the
      hadronization fractions of $\overline{b}$ (or $b$) quarks into
      the mesons indicated.  The inset plot shows (data-QCD)/QCD on a
      linear scale.  }}
\end{figure}

The signals for the two decay modes in Figure
\ref{fig:CDF_B_mass_plots} are combined to measure an ``average $B$
meson'' differential cross section, shown in Figure
\ref{fig:CDF_B_xsec}. The measured differential cross section is
consistent with the upper range of the theoretical prediction.  Thus,
this result shows the the same qualitative effect as the cross section
extracted from the inclusive lepton method.  Integrated over the range
$P_T > 6$ GeV and for $-1 < y < 1$, the average $B$ meson cross
section is $2.4\pm0.3\pm0.4$ $\mu$b.

\begin{figure}[hbtp] 
  \resizebox{\textwidth}{!}{\includegraphics*[103,220][520,672]
    {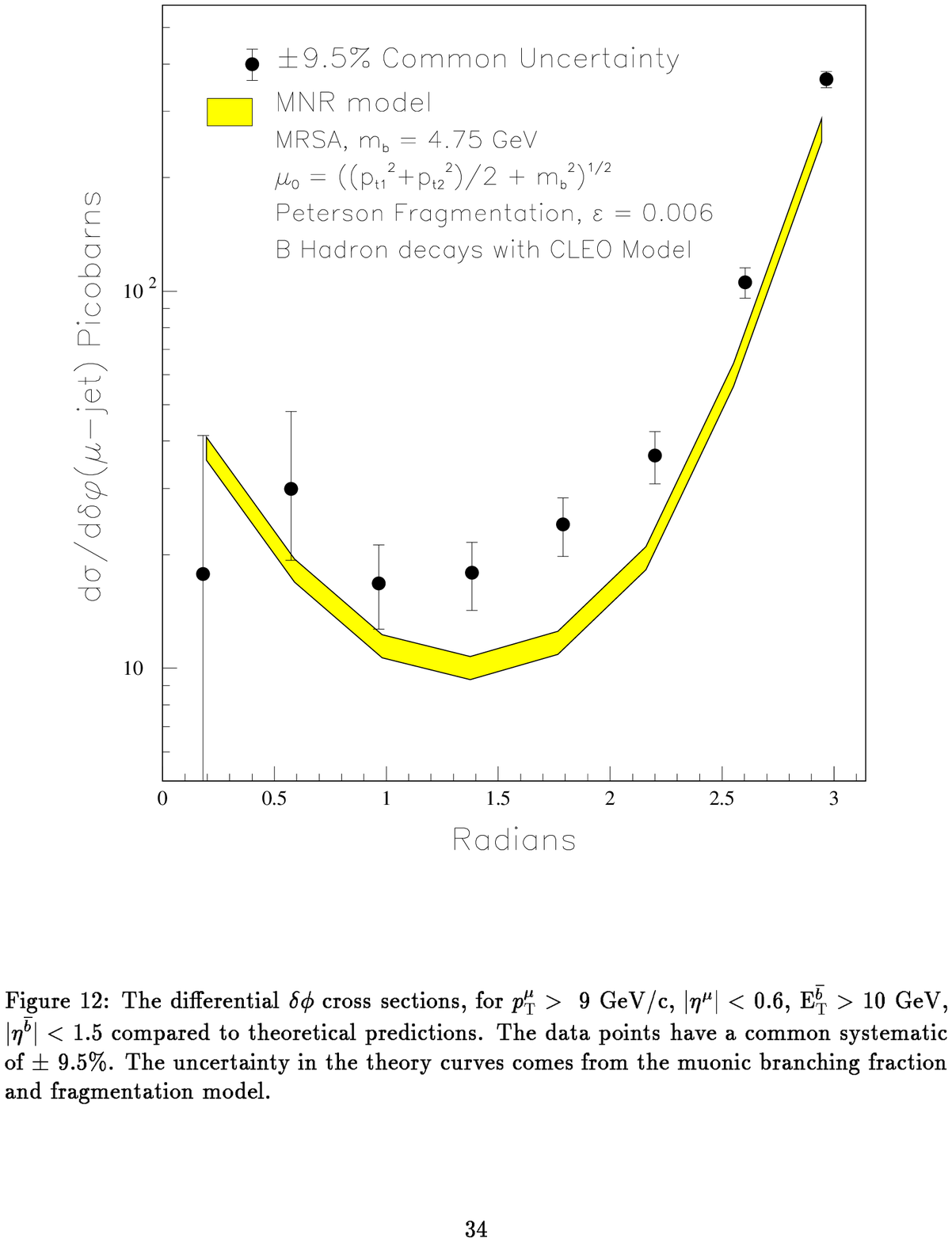} } {\caption{\label{fig:CDF_delta_phi} \protect
      \small (from ref. 9) The differential cross section for
      $\delta\phi$ from the CDF experiment.  The angle $\delta\phi$ is
      measured in the transverse plane between the directions of the
      muon from a $b$ quark and the jet from the $\overline{b}$ quark.
      The $\overline{b}$ jet must have a vertex tag, $E_T > 10$ GeV
      and $|\eta|<1.5$ . The muon must satisfy $p_T > 9$ GeV and
      $|\eta|<0.6$. The data points have a common systematic error of
      $\pm9.5\%$. The uncertainty in the theoretical curves comes from the
      muonic branching fraction and fragmentation model. }}
\end{figure}

Yet another type of measurement is available as a check on the
inclusive lepton and exclusive decay methods discussed above.  This
(third) method is the measurement of the associated $b\overline{b}$
pair where the muon is tagged from one $b$ quark and the jet is tagged
from the other. This process has a distinct signature in the plane
transverse to the line of collision. The $b$ and $\overline{b}$ quarks
emerge nearly back to back, the deviation from collinearity being due
to the transverse momentum of the initial state partons, for example,
the gluon pair.  When the $b$ quark fragments, the muon tends to
retain the direction of the $b$ quark. Thus, the muon and jet from the
accompanying $\overline{b}$ quark have a relative angular
distribution, $\delta\phi$, in the transverse plane that peaks at
$\pi$ radians.  The differential cross section for $\delta\phi$ from
the CDF experiment \cite{cdf_delta_phi} is shown in Figure
\ref{fig:CDF_delta_phi}.  The shape of the cross section is in
reasonable agreement with the QCD prediction shown, but the data lie
above the QCD prediction in absolute value.  Thus it appears from a
variety of measurements that the available QCD predictions
underestimate the absolute $b\overline{b}$ cross section while being
in good agreement with the shape of differential cross sections.

\begin{figure}[hbtp] 
%  \resizebox{\textwidth}{!}{\includegraphics*[70,282][563,561]
%    {CDF_Lambda_b_mass_plot.ps} }
%Line below for scanned image.
  \resizebox{\textwidth}{!}{\includegraphics*[0,221][595,561]
    {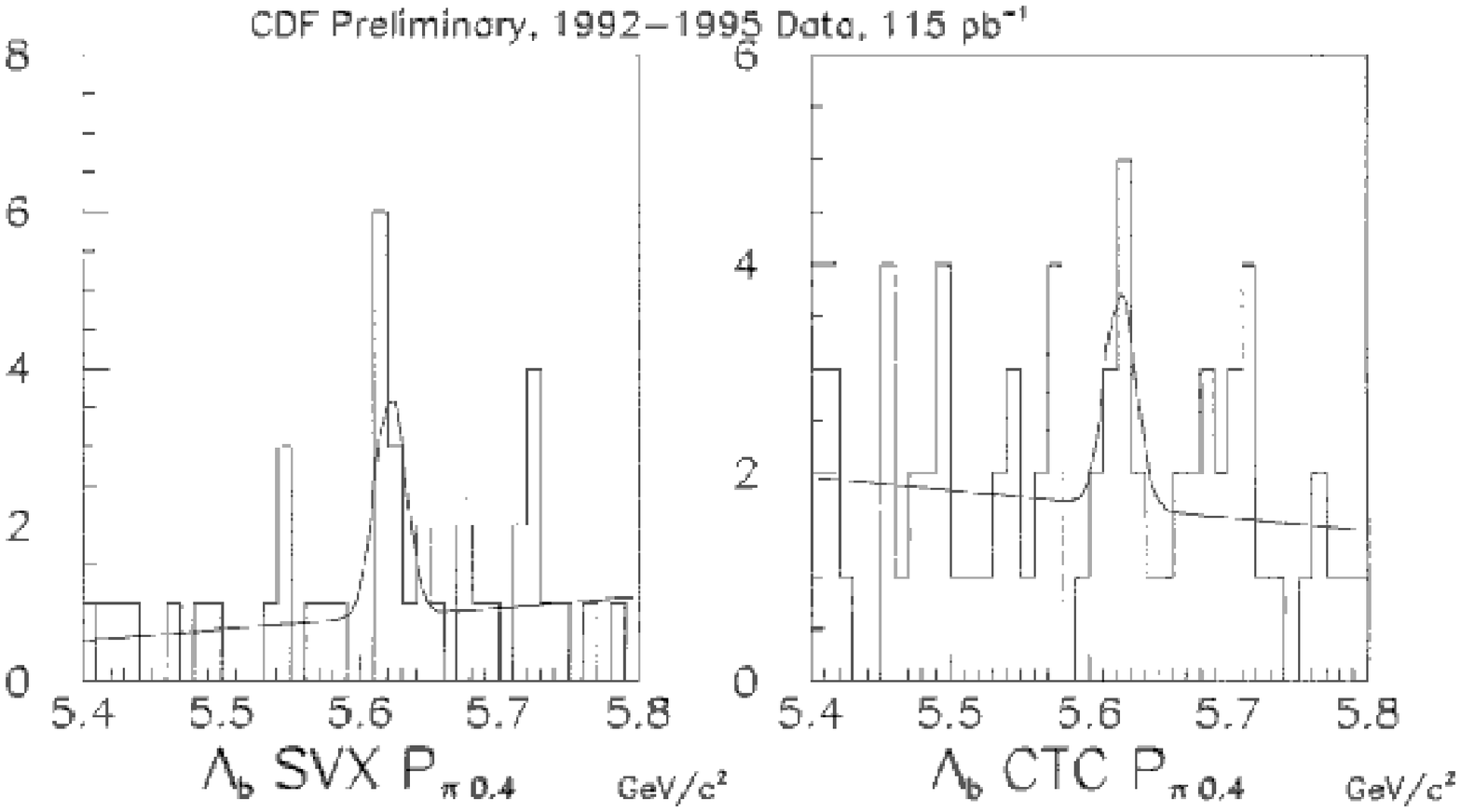}}
%  \vspace{5cm}
  {\caption{\label{fig:CDF_Lambda_b_mass_plot} \protect \small (from
      ref. 9) Invariant mass for $\Lambda_b \rightarrow J/\psi
      \Lambda$, $ J/\psi \rightarrow \mu^+\mu^-$, $\Lambda \rightarrow
      p \pi^-$.  The vertical scale is number of events per 10 MeV.
      The cuts are discussed in the text.  The plot on the left is for
      muons with vertex detector hits; that on the right for muons
      without vertex hits.  }}
\end{figure}

Measurement of the total beauty cross section requires measurements
for both beauty mesons and baryons. Information on $B$ baryon
production is just now becoming available.  The CDF collaboration
\cite{CDF_web_page} has studied the decay chain
\begin{equation}
\Lambda_b \rightarrow J/\psi ~\Lambda, 
 ~~J/\psi \rightarrow \mu^+\mu^-, 
 ~~\Lambda \rightarrow p \pi^-. 
\end{equation}
The charged tracks are required to have $p_T > 0.4$ GeV. Vertex
requirements are placed on the $\Lambda$,$ J/\psi$, and $\Lambda_b$.
Cuts are also placed on the reconstructed masses of the $\Lambda$ and
$ J/\psi$.  A signal of $8\pm4$ events is visible in the invariant
mass plot for $\Lambda_b$ shown in Figure
\ref{fig:CDF_Lambda_b_mass_plot}, on the left, for muon tracks
measured with vertex detector hits.

The cross section for $\Lambda_b$ can be measured, under certain
assumptions, by comparing the measured signal for $\Lambda_b
\rightarrow J/\psi \Lambda$ with the signal for $B_d \rightarrow
J/\psi K^0_s$. Both signals have similar topologies in the detector.
From the signal sizes and acceptances it is found that,
\begin{equation}
{\sigma \cdot Br(\Lambda_b \rightarrow J/\psi \Lambda) \over
\sigma \cdot Br(B_d \rightarrow J/\psi K^0_s)} = 0.31\pm0.15\pm0.06.
\end{equation}
{\em Assuming} that,
\begin{equation} 
\sigma(\Lambda_b) / \sigma(B_d) = 0.1 /0.375
\end{equation}
and using, 
\begin{equation}
Br(B_d \rightarrow J/\psi K^0_s)=3.7 \times 10^{-4}, 
\end{equation}
which is consistent with the measured value, gives, 
\begin{equation}
Br(\Lambda_b \rightarrow J/\psi \Lambda) =4.3\pm2.1\pm0.8 \times 10^{-4}.  
\end{equation}
To the extent that we consider the later $Br$ to be reasonable, the
assumed ratio of cross sections is also reasonable and hence provides
a rough estimate of the $\Lambda_b$ cross section.

\section{Quarkonia}

\subsection{Theory}

The `quarkonia' are bound states of a heavy quark and its anti- quark.
Charmonium is a $c\overline{c}$ state and bottomonium is a 
$b\overline{b}$ state.  The states
are further characterized by their 
spin, orbital angular momentum and radial quantum numbers.
The $L=0$ charmonium states are the $J/\psi$ and $\psi$ family;
the $L=0$ bottomonium states are the $\Upsilon$ family. These $L=0$ states
have quantum numbers $J^{PC} = 1^{--}$.

\begin{figure}[hbtp] 
%  \vspace{3in} 
\begin{center}
  \resizebox{!}{12cm}{\includegraphics*[183,327][396,741]
    {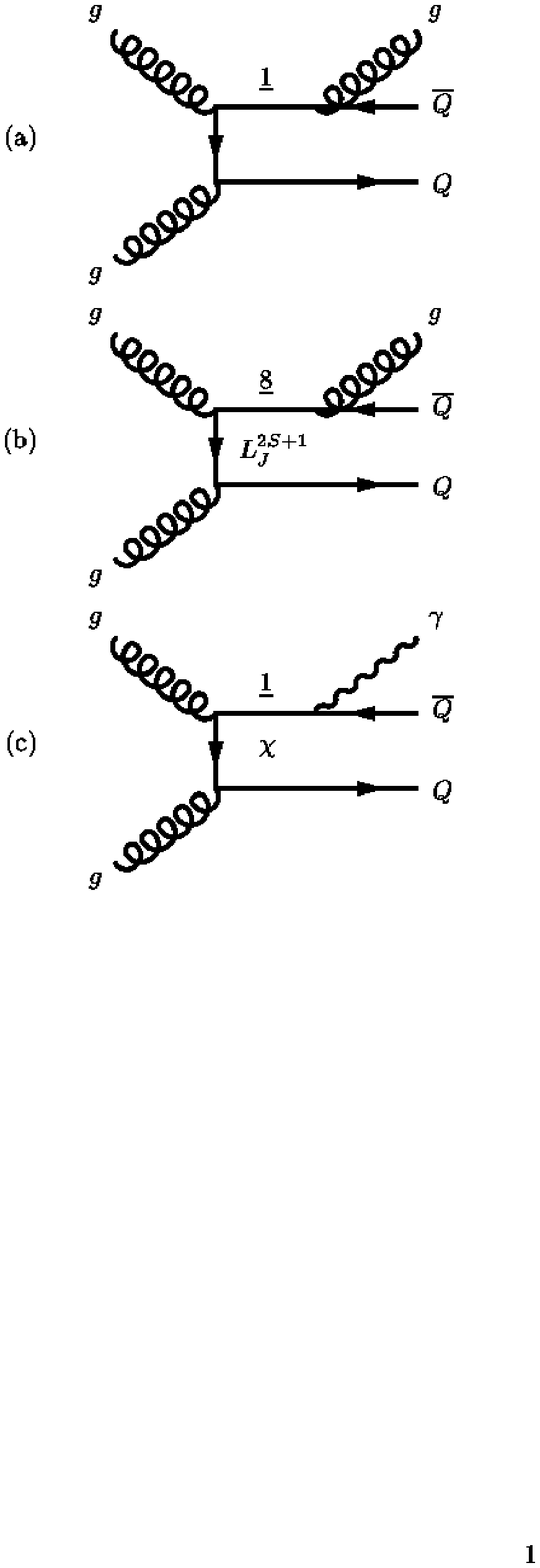} }
{\caption{\label{fig:onium_diagrams} \protect \small
      Feynman diagrams illustrating two gluon production of $J^{PC} =
      1^{--}$ quarkonia. (a) Direct production via a color singlet
      intermediate state with gluon emission. (b) Direct production
      via a color octet intermediate state with gluon emission. (c)
      Direct production of $\chi$ and subsequent decay to $\gamma$ +
      $1^{--}$ quarkonium.  }}
\end{center}
\end{figure}

The theory of the 
hadroproduction of quarkonia is discussed in recent papers by
Braaten and Fleming \cite{Braaten} and Cho and Leibovich \cite{Cho}.
The odd charge conjugation quantum number of the  $1^{--}$ states leads to
a fundamental restriction on their production in gluon collisions. The simple
process $gg \rightarrow Q\overline{Q}(1^{--})$ is not allowed since gluons
have $C=-1$.  However, a similar production 
process is allowed in which a gluon is
radiated from a final state quark line, as shown in Figure 
\ref{fig:onium_diagrams}(a).  This diagram is of order $\alpha_S^3$.
If the intermediate state is a color singlet $\underline{1}$, this process is
called direct color singlet production. But, the intermediate state 
need not be restricted to  a color singlet.
 
The intermediate state may be a either a color singlet or 
octet $\underline{8}$. Furthermore, the intermediate state can be have 
the full range of quantum numbers $^{2S+1}L_J$.  Thus, the general process,
\begin{equation}
gg \rightarrow Q\overline{Q}(^{2S+1}L_J) \rightarrow 
g + Q\overline{Q}(1^{--})
\end{equation}
is possible for the production of the $J/\psi$ and $\Upsilon$ families.
Examples of the intermediate state are $^3S_1$, $ ^1S_0$, and $^3P_J$. 
An example of direct color octet production is shown in the diagram of
Figure \ref{fig:onium_diagrams}(b).

A third important process, also of order $\alpha_S^3$, involves the direct 
production of the $\chi$ states which are characterized by $L=1$ and the
quantum numbers $PC=++$. Since $C=+$, $\chi$ can be produced directly
from a two gluon collision, without the emission of a final state gluon,
as shown in Figure \ref{fig:onium_diagrams}(c).  

The $\chi_{C1}$ ($J=1$) and  $\chi_{C2}$ ($J=2$) states
have large branching fractions into $J/\psi + \gamma $. 
Thus, these states can be studied experimentally  by detecting 
the $J/\psi$ alone or both final state particles.  However, the 
$\chi_{C0}$ ($J=0$) has a small branching fraction
into $J/\psi + \gamma$ and hence cannot be easily studied.

From inspection of the simple production diagrams considered above, it
is evident that $\chi$ production has the most straightforward
theoretical treatment. There is no soft gluon emission required in the
fundamental production process, and furthermore, there is no
fragmentation process as for a state with open charm or beauty. Direct
$J/\psi$ or $\Upsilon$ production is complicated by the emission of
the soft gluon required by conservation of charge conjugation parity
of the strong interaction.

\subsection{Charmonium}

The $J/\psi$ was co-discovered in a hadronic fixed target experiment
and has since been studied in that way with high statistics. A number of
results have been reported recently from Fermilab fixed target experiments
using high energy 
$\pi^-$ \cite{E672} and proton \cite{E771,E789} beams. As an
illustration, we consider the measurement of $J/\psi$ and $\psi(2S)$ production
from Fermilab experiment E789 which used the extracted 800 GeV proton beam.

\begin{figure}[hbtp] 
  \resizebox{\textwidth}{!}{\includegraphics*[20,144][537,664]
    {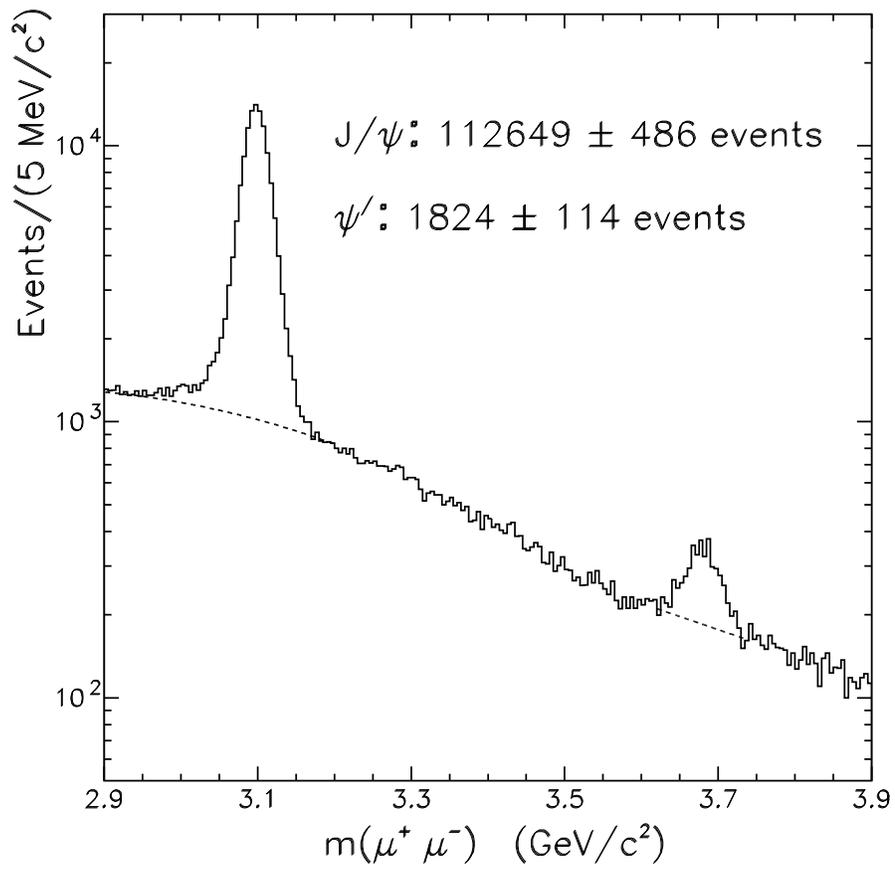} }
  {\caption{\label{fig:E789_JPsi_mass_plot} \protect \small (from ref.
      15) Invariant mass of $\mu^+\mu^-$ pairs from Fermilab E789. The
      dashed curves show the fits to the combinatoric background.  The
      name $\psi'$ is in common usage for the $\psi(2S)$ }}
\end{figure}

The $\mu^+\mu^-$ invariant mass plot from E789 is shown in Figure
\ref{fig:E789_JPsi_mass_plot}.  Of particular note is the large number
of signal events allowing for detailed studies of the $x_F$ and $P_T$
dependence of the production cross section.

\begin{figure}[hbtp] 
  \resizebox{\textwidth}{!}{\includegraphics*[20,144][537,664]
    {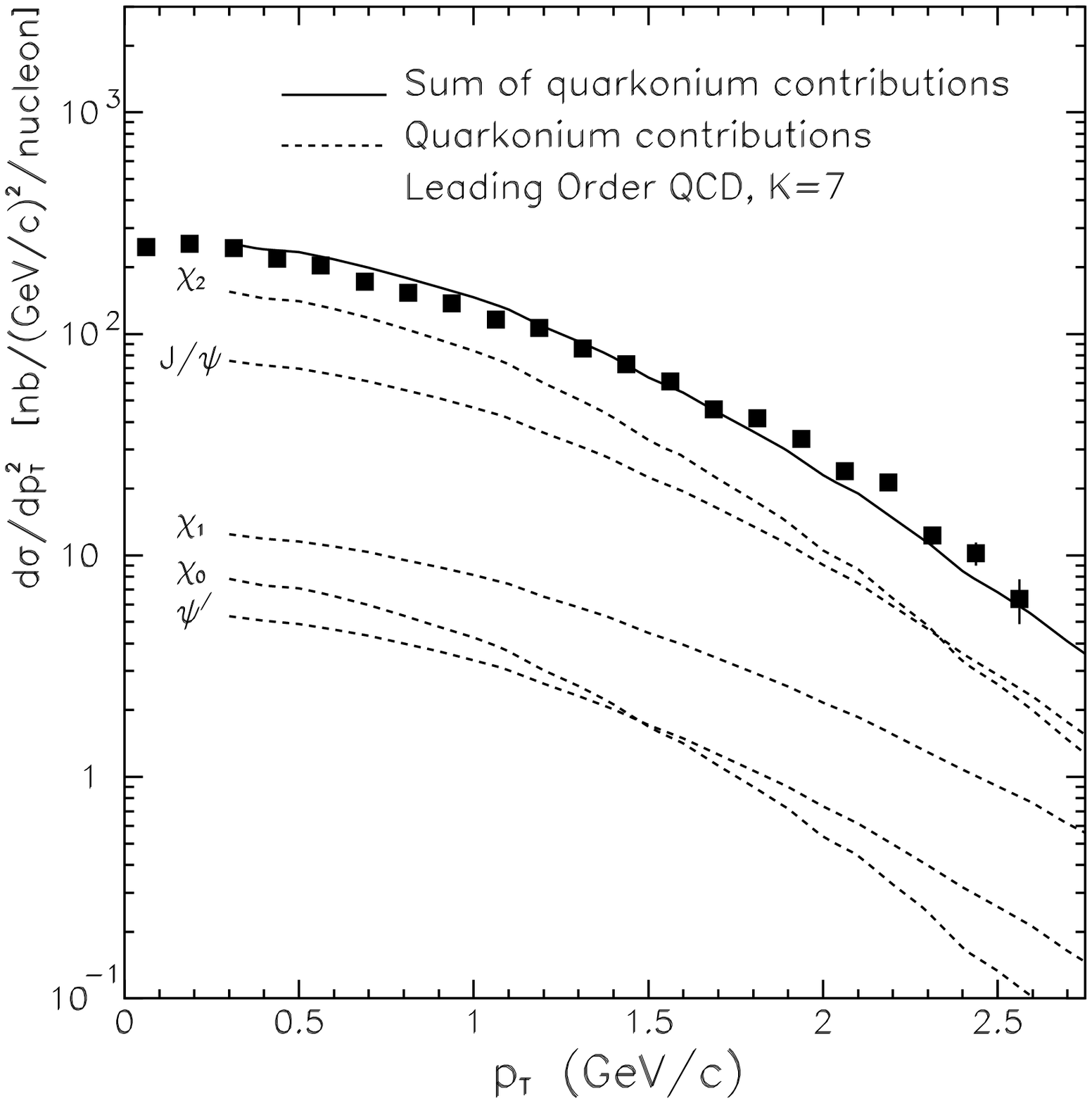} } {\caption{\label{fig:E789_JPsi_Pt_dist}
      \protect \small (from ref. 15) Measurement from Fermilab E789 of
      the differential cross section for $J/\psi$ production versus
      the transverse momentum of the $J/\psi$.  The 20\% systematic
      normalization uncertainty is not shown.  The dashed curves are 7
      times the leading-order predictions for the inclusive $P_T$
      distributions of $J/\psi$ mesons originating from various
      quarkonium states.  The solid curve is the sum of the
      quarkonium contributions. (The contribution from $B \rightarrow
      J/\psi + X$ decays is negligible.)  }}
\end{figure}

The differential cross section in $P_T$ for $J/\psi$ 
is shown in Figure 
\ref{fig:E789_JPsi_Pt_dist}.  
Also shown are predictions from a computer program of Mangano based on
leading order QCD.  Shown separately are the contributions to 
the cross section from direct production and from the decay of $\chi_C$
states. The sum of the contributions seems to fit the data well.
But actually the curves shown are 7 times the prediction! Thus, the theory
predicts the shape reasonably well but severely underestimates the absolute
scale of the cross section. This situation is even more dramatic for
$\psi(2S)$ production.

\begin{figure}[hbtp] 
  \resizebox{\textwidth}{!}{\includegraphics*[20,144][537,664]
    {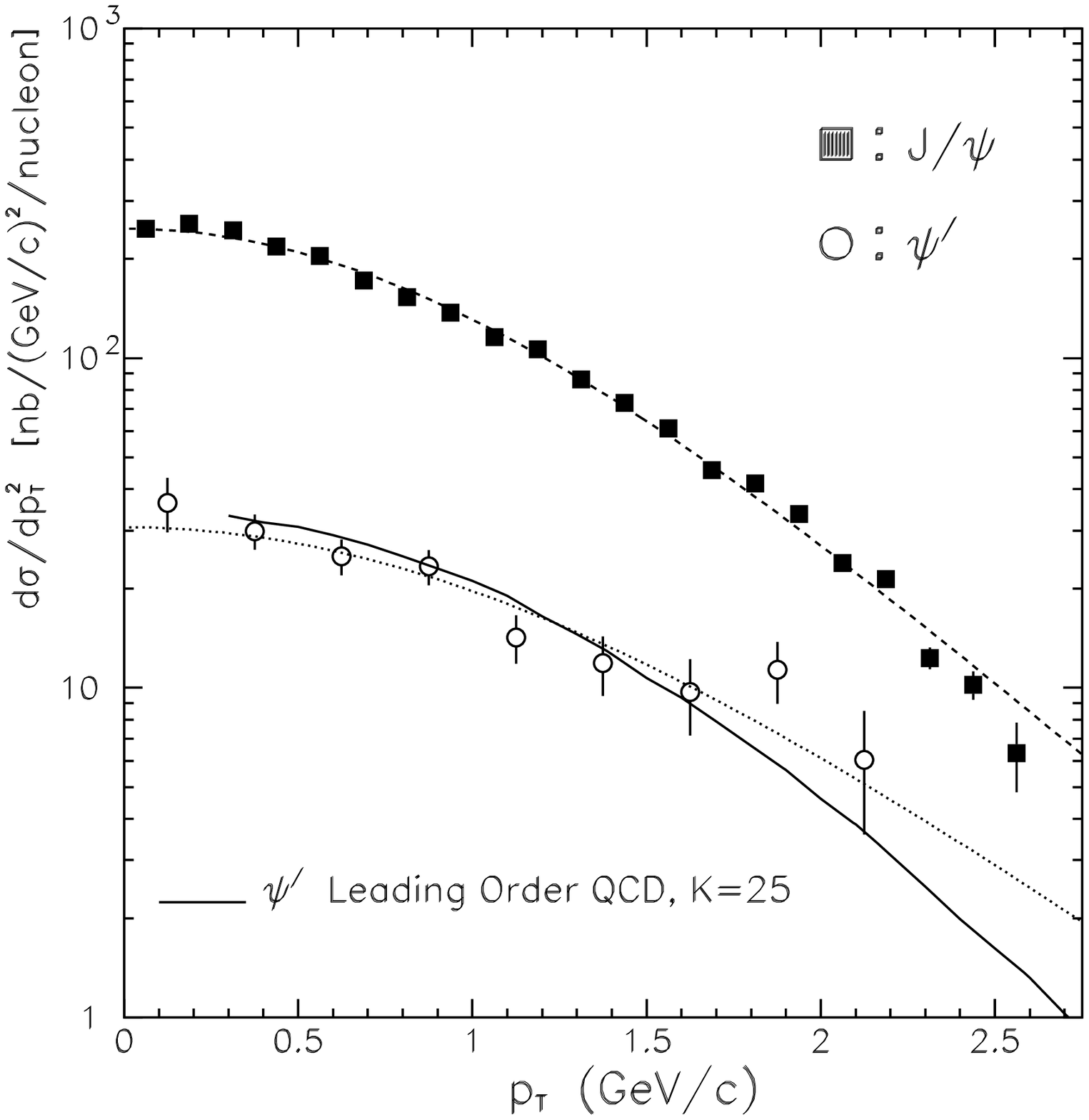} }
  {\caption{\label{fig:E789_Psi2s_Pt2_dist} \protect \small (from ref.
      15) E789 measurement of $d\sigma/dP_T^2$ for $pN \rightarrow
      \psi(2S)+ X$ versus $P_T$ of the $\psi(2S)$.  Also shown is the
      measurement of $J/\psi$ for comparison.  The systematic
      normalization uncertainty of 30\% for the $\psi(2S)$ is not
      shown.  The solid curve is 25 times the leading order prediction
      for the $\psi(2S)$. }}
\end{figure}

The E789 measurement of $d\sigma/dP_T^2$ for $\psi(2S)$ is shown in Figure
\ref{fig:E789_Psi2s_Pt2_dist}. A multiplicative factor of 25 is required 
for the leading order prediction to match 
the scale of the data.  Like the $J/\psi$, the
shape of the prediction for $\psi(2S)$ 
is in reasonable agreement with the data.

\begin{figure}[hbtp] 
%  \vspace{8cm}
% Line below for scanned image.
  \resizebox{\textwidth}{!}{\includegraphics*[15,194][586,560]
    {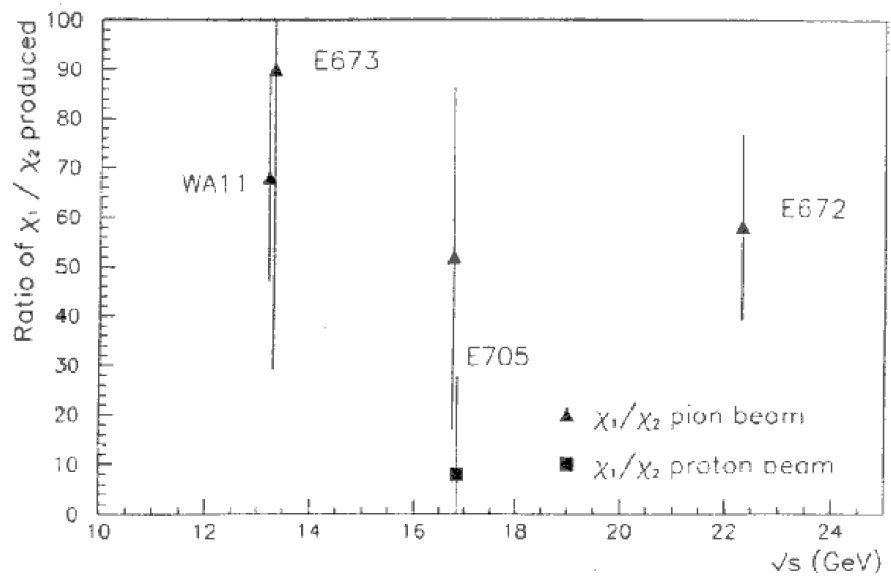}} {\caption{\label{fig:chi_ratios} \protect
      \small (from ref. 15) Measurements of $\sigma(\chi_{C1}) /
      \sigma(\chi_{C2})$, in \%, from fixed target experiments.  The
      prediction from spin counting is 0.6.  }}
\end{figure}

As mentioned in the theory section, measurements of $\chi_C$ production are 
important because of the firmer theoretical understanding of $\chi$ versus
vector meson quarkonium production. Unfortunately, the measurement of the
photon is typically inefficient in detecting the decay 
$\chi_C \rightarrow J\psi + \gamma$.  A review of fixed target data
on $\chi_C$ production is given by McManus \cite{McManus}. One simple 
prediction is that the ratio of cross sections for $\chi_{C1}$ and 
$\chi_{C2}$ is given by spin counting.  Thus, one expects,
\begin{equation}
{ \sigma(\chi_{C1}) \over \sigma(\chi_{C2}) } = 
{ 2(1) + 1 \over 2(2) + 1 } = 0.6 
\end{equation}
Data for this ratio from various experiments are plotted in Figure
\ref{fig:chi_ratios}.  Within the large statistical errors, the data
are indeed consistent with the simple spin counting prediction.  This
result is consistent with the picture that 
leading order QCD processes are sufficient
to explain $\chi$ production.

\begin{figure}[hbtp] 
  \resizebox{\textwidth}{!}{\includegraphics*[323,514][551,728]
    {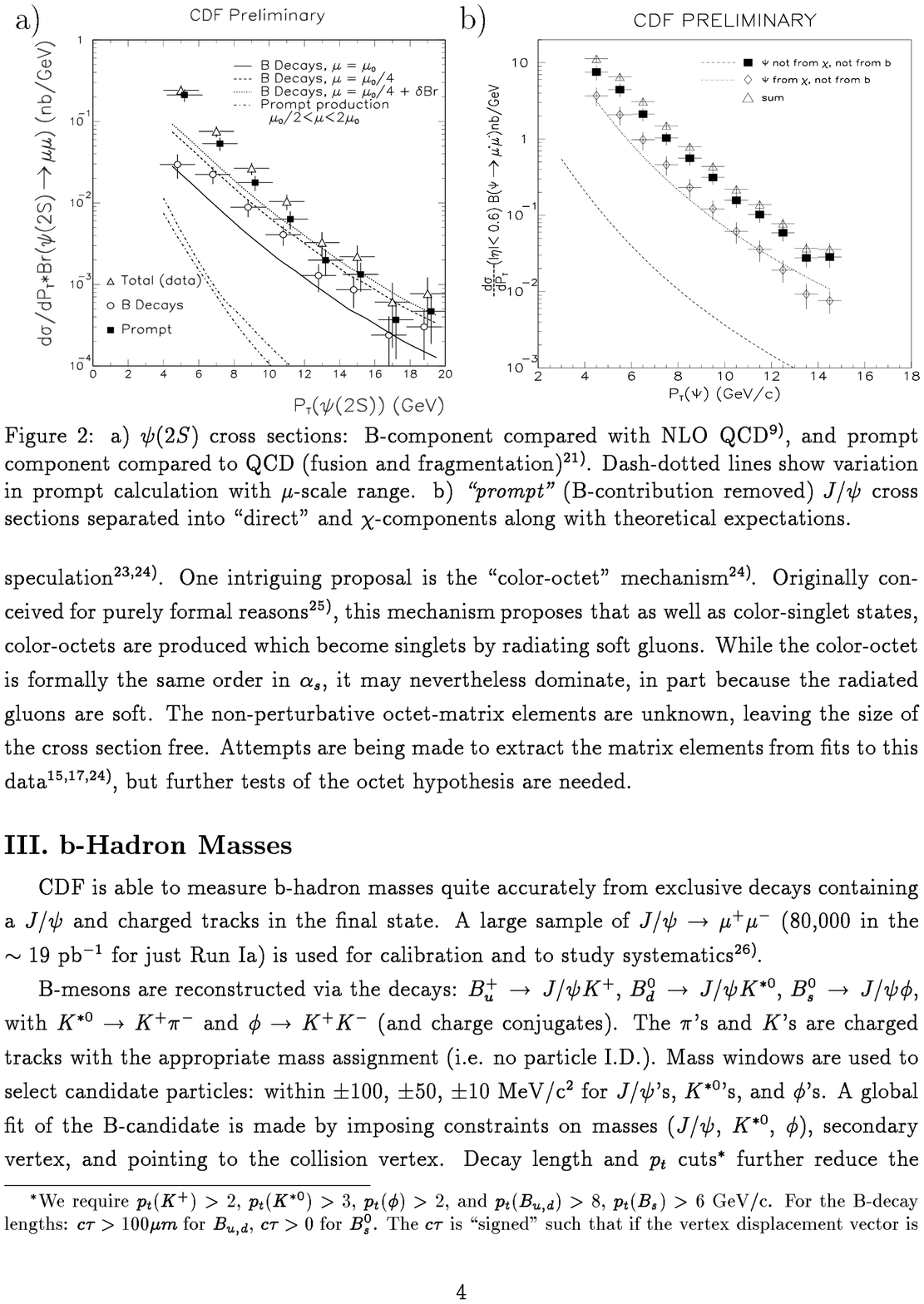} } {\caption{\label{fig:CDF_JPsi}
      \protect \small (from ref. 17) CDF measurement of $J/\psi$
      production. Data is shown for the cross section times branching
      fraction for $p\overline{p} \rightarrow J/\psi + X$, $J/\psi
      \rightarrow \mu^+\mu^-$ versus $P_T$ of the $J/\psi$.  The
      contribution of $J/\psi$ from $B$ decay is eliminated using
      vertex information.  The contribution of $J/\psi$ from $\chi_C$
      decays is measured separately by explicit reconstruction of the
      decay $\chi_C \rightarrow J/\psi + \gamma$.  The measured cross
      section for prompt $J/\psi$ production (including the small
      contribution from $\chi_C$ decays) is given by the open
      triangles. The perturbative QCD prediction, which is dominated
      by the contribution from $\chi_C$ production, underestimates the
      data by a factor between 5 and 10 over the $P_T$ range measured.
      }}
\end{figure}

There have also been recent measurements of charmonium production from
the Tevatron Collider. The CDF collaboration has reported preliminary
measurements \cite{Bauer} of the differential cross sections for
$J/\psi$ and $\psi(2S)$ production. The data for the cross section
times branching fraction for $p\overline{p} \rightarrow J/\psi + X$,
$J/\psi \rightarrow \mu^+\mu^-$ versus $P_T$ of the $J/\psi$ is shown
in Figure \ref{fig:CDF_JPsi}.  The contribution of $J/\psi$ from $B$
decay is eliminated using vertex information.  The contribution of
$J/\psi$ from $\chi_C$ decays is measured separately by explicit
reconstruction of the decay $\chi_C \rightarrow J/\psi + \gamma$.  The
measured cross section for prompt $J/\psi$ production (including the
small contribution from $\chi_C$ decays) is given by the open
triangles. The perturbative QCD prediction, which is dominated by the
contribution from $\chi_C$ production, underestimates the data by a
factor between 5 and 10 over the $P_T$ range measured.  Thus, the
situation at E$_{CM}$ = 1.8 TeV is remarkably similar to that of the
fixed target data at E$_{CM}$ = 0.04 TeV. Note also that the CDF
measurement of $J/\psi$ from $\chi_C$ decay is in reasonable agreement
with the QCD prediction.

\begin{figure}[hbtp] 
%  \resizebox{\textwidth}{!}{\includegraphics*[83,508][305,728]
%    {CDF_Psi_and_Psi2S_Pt_dist.ps} }
% Lines below for scanned image.
  \resizebox{\textwidth}{!}{\includegraphics*[15,95][595,670]
    {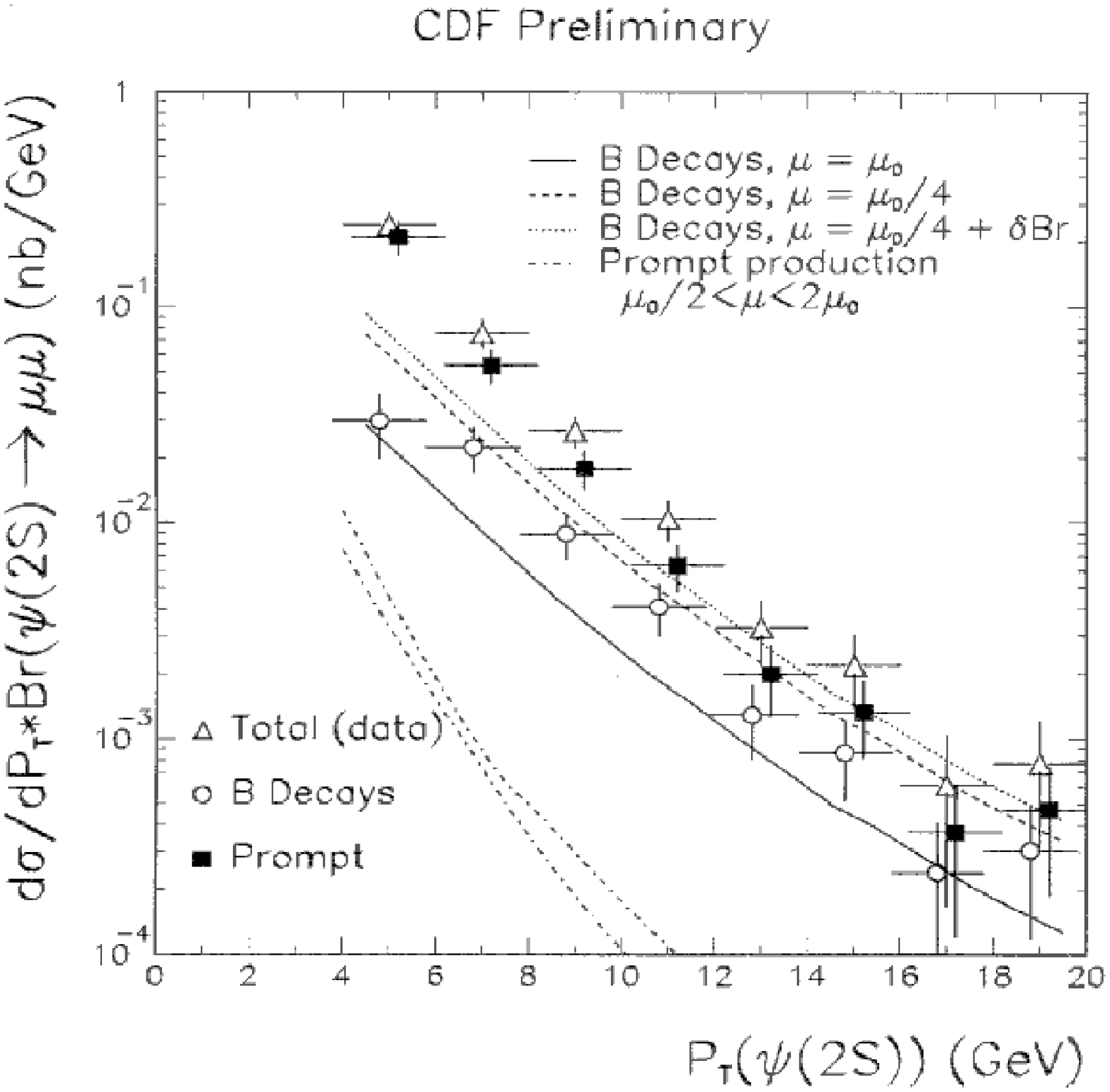} } 
      {\caption{\label{fig:CDF_psi2s} \protect
      \small (from ref. 17) CDF measurement of $\psi(2S)$ production.
      The prompt $\psi(2S)$ data points are the solid squares.  The
      QCD prediction for prompt production is the area between the
      dash-dotted lines.  The data and theory disagree in both shape
      and normalization. }}
\end{figure}

The CDF measurements for $\psi(2S)$ production are shown in Figure 
\ref{fig:CDF_psi2s}.  The prompt $\psi(2S)$ data points are the solid
squares.  The QCD prediction for prompt production is the area between the
dash-dotted lines.  The data and theory disagree in both shape and 
normalization.  At the lowest $P_T$ the prediction 
is at least 20 times too low. As $P_T$ increases, the discrepancy 
gets even larger.  These results, along with those at fixed target energies,
indicate that non-perturbative processes (such as the soft gluon emission
shown in Figure \ref{fig:onium_diagrams}) 
play an important role in prompt $J/\psi$ and $\psi(2S)$ production.

\begin{figure}[hbtp] 
  \resizebox{\textwidth}{!}{\includegraphics*[182,270][411,498]
    {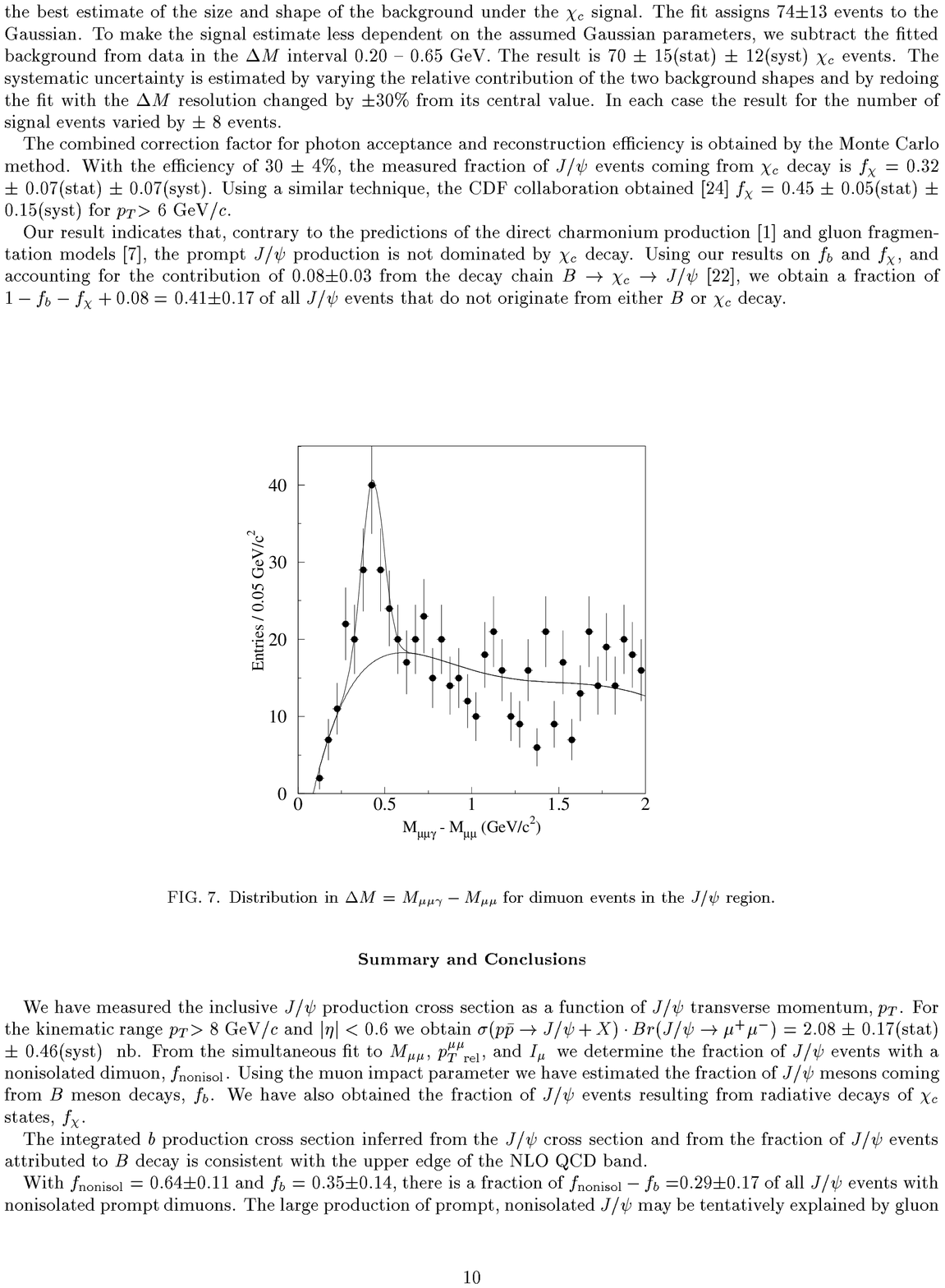} } {\caption{\label{fig:D0_Chi_c_mass_plot}
      \protect \small (from ref. 18) Signal for $\chi_C$ from the D0
      experiment.  Shown is the distribution in $\Delta M =
      M_{\mu\mu\gamma} - M_{\mu\mu}$ for dimuon events in the $J/\psi$
      region.  The dimuon pair must satisfy $P_T(\mu\mu) > 8$ GeV.  }}
\end{figure}

Explicit reconstruction of $\chi_C$ has also been accomplished with
the D0 experiment \cite{D0_chiC} at the Fermilab Tevatron. The signal
can be seen in Figure \ref{fig:D0_Chi_c_mass_plot} which shows the
distribution of the difference in invariant mass between the
reconstructed $J/\psi \rightarrow \mu^+\mu^-$ and the candidate
$\chi_C \rightarrow \mu^+\mu^-\gamma$ combinations.  Note that the
resolution is not sufficient to resolve the $J=0,1,2$ $\chi_C$ states
as is possible in fixed target experiments.  It is found in D0 that
the fraction $f_{\chi}$ of $J/\psi$'s originating from $\chi_C$ is,
\begin{equation}
f_{\chi} = 0.32\pm0.07\pm0.07.
\end{equation}
That this ratio is far from one indicates that the bulk of $J/\psi$ production
is direct and not from $\chi_C$ decay.  This reinforces the importance of
the non-perturbative process of direct production.

\subsection{$\Upsilon$ Production}

\begin{figure}[hbtp] 
  \resizebox{\textwidth}{!}{\includegraphics*[80,199][537,633]
    {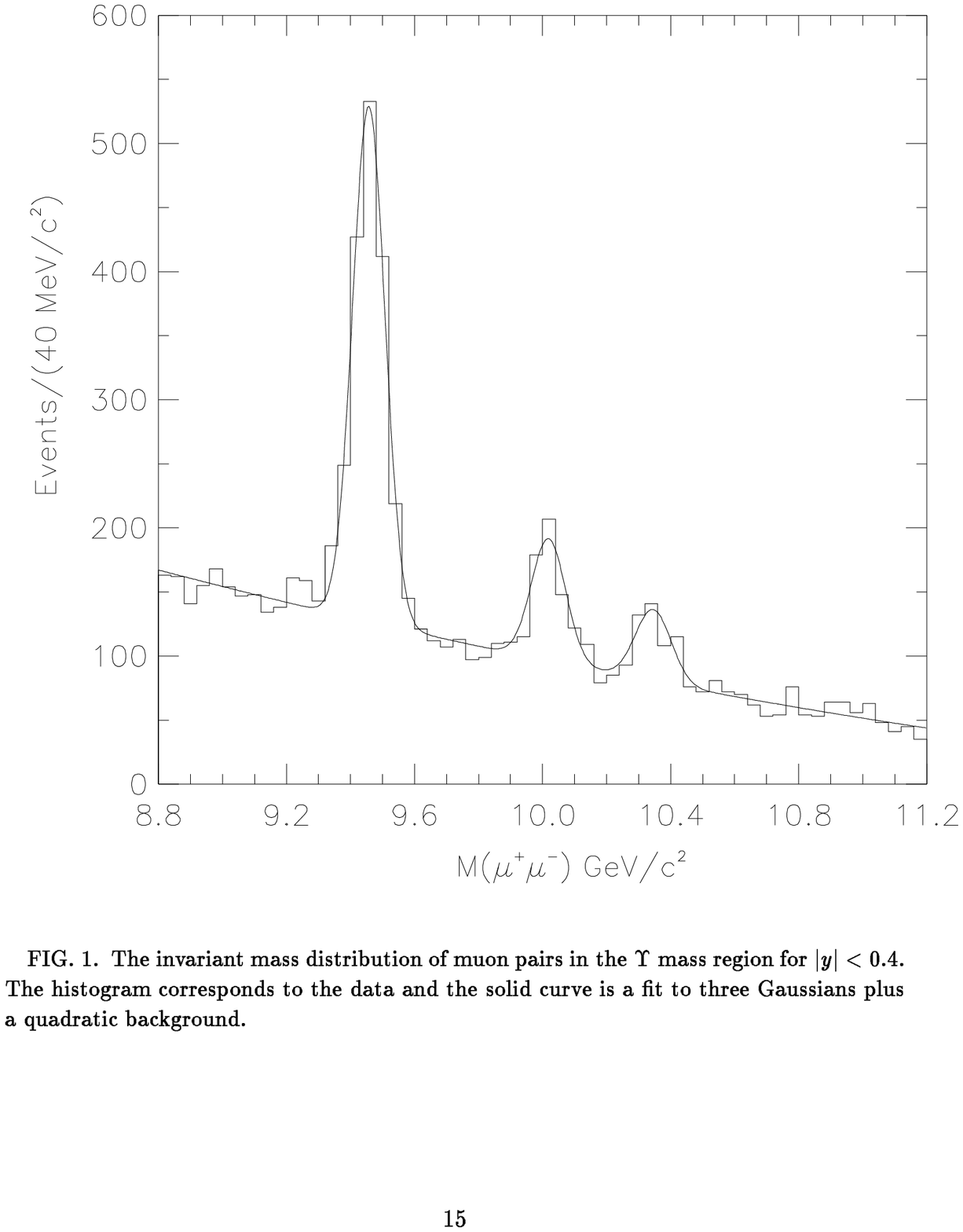} } {\caption{\label{fig:cdf_up_mass_plot}
      \protect \small (from ref. 19) CDF measurement of the invariant
      mass distribution of muon pairs in the $\Upsilon$ mass region
      for $|y|<0.4$.  There are also cuts on the $P_T$ of the muons.
      The histogram corresponds to the data and the solid curve is a
      fit to three Gaussians plus a quadratic background.  }}
\end{figure}

\begin{figure}[hbtp] 
  \resizebox{\textwidth}{!}{\includegraphics*[105,374][482,749]
    {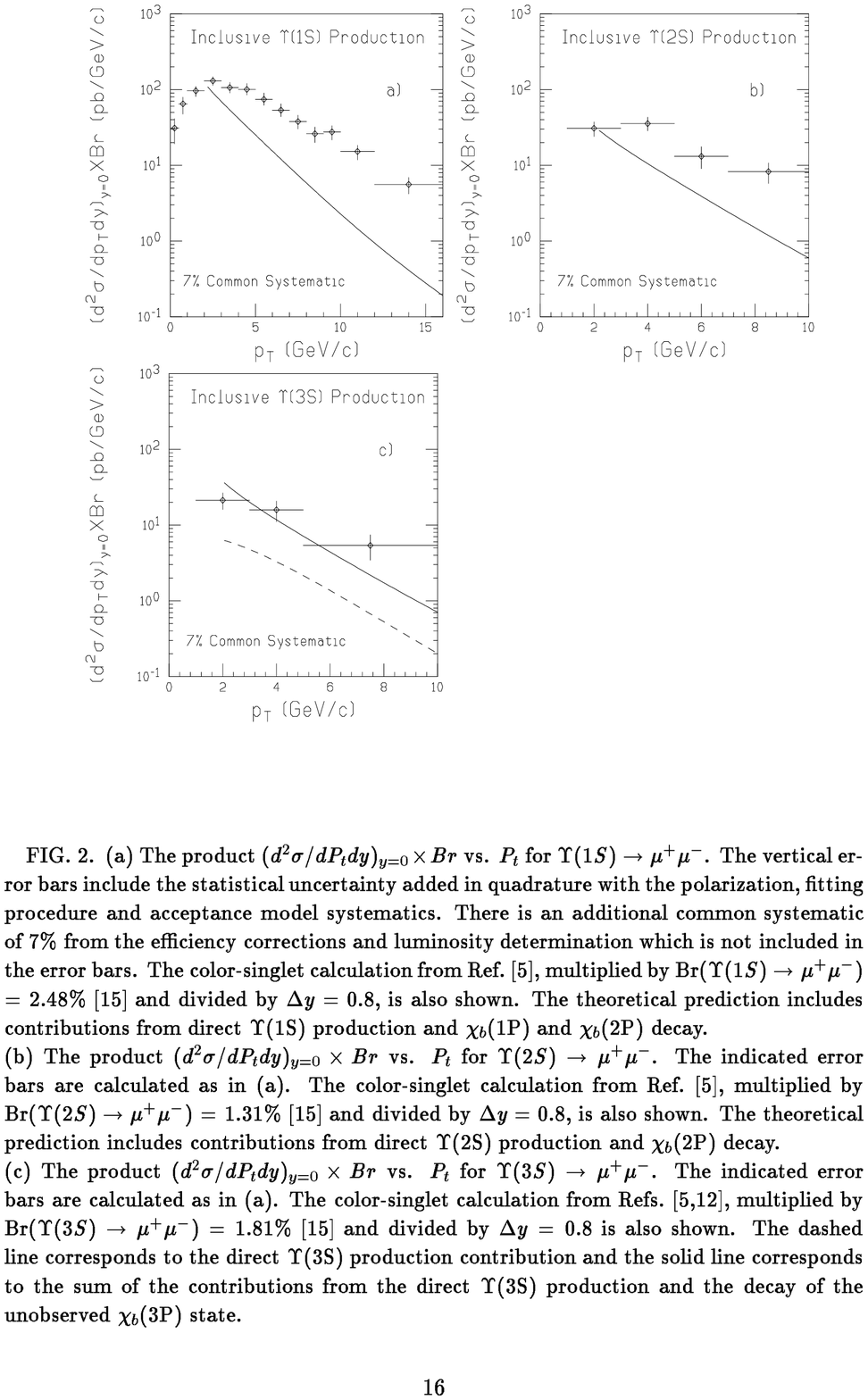} }
  {\caption{\label{fig:CDF_Upsilon_Pt_dist} \protect \small (from ref.
      19) CDF data for $\Upsilon$ production compared to the
      predictions of the QCD color singlet model of Cho and
      Leibovitch.  Plotted is the product of the cross section times
      branching fraction $(d^2\sigma/dP_Tdy)_{y=0} \times Br $ versus
      $P_T$ for $\Upsilon$.  The predictions (solid lines) include
      direct production and feeddown from $\chi_B$. In c), the dashed
      line corresponds to direct production only, whereas the solid
      line also includes the contribution from the decay of the
      unobserved $\chi_B(3P)$ state. }}
\end{figure}

The $\Upsilon$ was discovered using fixed target hadroproduction and
has now been well measured at the Fermilab Tevatron collider.  An
invariant mass spectrum for $\Upsilon \rightarrow \mu^+\mu^-$ from
CDF\cite{CDF_upsilon} is shown in Figure \ref{fig:cdf_up_mass_plot}.
The experimental resolution is good enough to see separate peaks for
the $\Upsilon$ $1S$, $2S$ and $3S$ radial states. The differential
cross sections in $P_T$ for these particles are shown in Figure
\ref{fig:CDF_Upsilon_Pt_dist}. Also shown are the predictions of the
color singlet model of Cho and Leibovitch.  The prediction includes
direct production and feeddown from $\chi_B$.  The predictions fall
below the data for the three particles shown. For the $\Upsilon(1S)$,
the shape of the prediction is not in agreement with the data. Thus,
it appears that the theoretical treatment of the production of the
$J=1$ bottomonium states suffers the same difficulty as for the $J=1$
charmonium states: nonperturbative effects are likely important.

\section{Concluding Remarks}

In this review, we've examined both open and closed charm and beauty
states produced at $CM$ energies spanning nearly two orders of
magnitude (0.02 TeV to 1.8 TeV).  Experimental accomplishments include
measurements of fully reconstructed particles yielding normalized
cross sections and their dependence on transverse and longitudinal
momentum.  For the open states, perturbative QCD calculations are
successful in explaining the shapes of the differential cross
sections but underestimate the scale of the cross sections by factors
of 2-3.  Furthermore, the theoretical predictions are too
sensitive to variations in the underlying parameters of the theory,
such as the scales of renormalization and factorization.  This leads
to theoretical uncertainties in the cross section scale of order of
the difference between the measurement and prediction.  

The picture emerging for quarkonia production is that for the $J/\psi$
and $\Upsilon$ families the production mechanism is fundamentally
non-perturbative.  This is evidenced by the large discrepancy between
the measured cross sections and those predicted using QCD perturbation
theory.  However, measurements of $\chi_C$ production seem to be
reasonably explained as due to perturbative processes.

Future experiments promise progress in several areas.  The Tevatron
collider experiments will continue to study production of open beauty,
and both charm and beauty quarkonia.  LHC experiments will open a new
energy regime.  A new fixed target experiment, SELEX, is underway at
Fermilab using $\Sigma$ and proton beams. The COMPASS experiment is
planned for the future CERN fixed target program using a hyperon beam.
The HERA-B experiment is under construction at DESY and will study
fixed target $pN$ production.

The challenge to theorists is to tame the theoretical errors in
perturbative QCD calculations with new approaches to factorization
and renormalization.  Models that can treat the fundamentally
non-perturbative processes in quarkonia production and in fragmentation
of heavy quarks would be an important tool to assess whether ``new''
physics is required.

The hadroproduction of charm and beauty particles serves as one of the
testing grounds for QCD.  Hadronic collisions also provide a copious
source of heavy quarks, allowing us to study their spectroscopy and
decay with great precision.  However, we must understand the
production characteristics well to achieve precision in the decay
measurements.  These incentives spark our interest in future work
in this field.

\section*{Acknowledgments}

The author appreciates helpful discussions with Prof. James Johnson of
Wayne State University.  Thanks to Prof. Kwong Lau of the University
of Houston for providing Figure \ref{fig:b_cross_ft}.  Prof. Stephen
Takach of Wayne State generously provided a careful reading
and critique of the manuscript.  The author is grateful for funding
from Wayne State and the United States Department of Energy.

\end{document}